\newif\ifreport\reporttrue

\ifreport
\documentclass[10pt, conference, letterpaper]{IEEEtran}
\else

\documentclass[10pt, conference, letterpaper]{IEEEtran}

\fi

\usepackage{amsfonts}
\usepackage{mathrsfs}
\usepackage{graphicx,epsfig,amssymb,amsmath,url,latexsym}
\makeatletter
\newif\if@restonecol
\makeatother

\usepackage[lined, boxed, linesnumbered, commentsnumbered, ruled]{algorithm2e} 
\usepackage{subfigure}
\usepackage[square,sort&compress,numbers]{natbib}
\usepackage{bm}
\usepackage{multirow}
\usepackage{color}
\usepackage{enumerate}
\usepackage{setspace}
\usepackage{graphicx}
\usepackage{bbm}
\usepackage{url}
\usepackage[utf8]{inputenc}
\usepackage[T1]{fontenc}

\usepackage{mathtools}

\DeclarePairedDelimiter{\ceil}{\lceil}{\rceil}
\DeclarePairedDelimiter\floor{\lfloor}{\rfloor}

\newtheorem{theorem}{Theorem}
\newtheorem{lemma}{Lemma}

\newtheorem{definition}{Definition}

\newtheorem{remark}{Remark}[theorem]

\IEEEoverridecommandlockouts

\begin{document}

\title{High Throughput Low Delay Wireless Multicast\\via Multi-Channel Moving Window Codes\thanks{This work was supported in part by NSF grants CNS-1719371, CNS-1446582, CNS-1518829, CNS-1409336, CNS-1547306, CNS-1302620, CNS-1514260, CNS-1254032, and ONR grants N00014-17-1-2417 and N00014-15-1-2166.}}


\author{Fei Wu$^\dag$, Yin Sun$^\S$, Lu Chen$^\dag$, Jackie Xu$^\dag$, Kannan Srinivasan$^\dag$, and Ness B. Shroff$^\dag{}^\ddag$ \\
	$^\dag$Dept. of CSE, $^\ddag$Dept. of ECE, The Ohio State University, Columbus, OH\\
	$^\S$Dept. of ECE, Auburn University, Auburn, AL\\}
	

\maketitle

\begin{abstract}

A fundamental challenge in wireless multicast has been how to simultaneously achieve high-throughput and low-delay for reliably serving a large number of users. In this paper, we show how to harness substantial throughput and delay gains by exploiting multi-channel resources. We develop a new scheme called Multi-Channel Moving Window Codes (MC-MWC) for multi-channel multi-session wireless multicast. The salient features of MC-MWC are three-fold. (i) High throughput: we show that MC-MWC achieves order-optimal throughput in the many-user many-channel asymptotic regime. Moreover, the number of channels required by a conventional channel-allocation based scheme is shown to be doubly-exponentially larger than that required by MC-MWC. (ii) Low delay: using large deviations theory, we show that the delay of MC-MWC decreases linearly with the number of channels, while the delay reduction of conventional schemes is no more than a finite constant. (iii) Low feedback overhead: the feedback overhead of MC-MWC is a constant that is independent of both the number of receivers in each session and the number of sessions in the network.
Finally, our trace-driven simulation and numerical results validate the analytical results and show that the implementation complexity of MC-MWC is low.
\end{abstract}

\section{Introduction}
\label{sec:introduction}
Mobile video is expected to contribute $75\%$ of all the mobile traffic by 2020 \cite{cisco}. Wireless multicast, by leveraging the shared nature of the wireless medium, could be an efficient way to distribute popular videos to many users. However, wireless multicast has not been widely deployed in practical systems, because it has been difficult to simultaneously achieve high throughput, low delay, and low feedback overhead for serving a large number of users.

\emph{Throughput challenge:} In a wireless communication system, the channel conditions of different receivers are heterogeneous due to multipath, shadowing, mobility, etc. Since the throughput of multicast is bottlenecked by the receiver with the worst channel condition, as the number of receivers grows, the achievable throughput in multicast vanishes, offsetting the multicast gain.

\emph{Delay challenge:} Existing multicast coding schemes incur large delay at the receivers. There are two categories of multicast coding schemes in the literature. The first category of schemes employ a block coding strategy, e.g., random linear network coding (RLNC) \cite{ho2006random}, LT codes \cite{luby2002}, and Raptor codes \cite{shokrollahi2006raptor}, etc. However, to maintain a non-diminishing throughput, the block size has to grow on the order\footnote{We use the standard order notation: for two real-valued sequences $\{x_n\}$ and $\{y_n\}$, $x_n=o(y_n)$ if $\lim_{n\to\infty}x_n/y_n=0$; and $x_n=\omega(y_n)$ if $\lim_{n\to\infty}x_n/y_n=\infty$; and $x_n=\Omega(y_n)$ if $\lim_{n\to\infty}x_n/y_n\ge z$ for some constant $z>0$; and $x_n=\Theta(y_n)$ if $z_1\le\lim_{n\to\infty}x_n/y_n\le z_2$ for some constants $z_1>0$ and $z_2>0$.} $\Omega(\log n)$ as the number of receivers $n$ grows \cite{yang2012throughput}. Since the decoding delay increases with the block length, such schemes are known to have a poor delay performance. The second category of schemes achieve a lower delay through an incremental network coding design, e.g., \cite{MWNC,kumar2008arq,sundararajan2009feedback,parastoo2010optimal,Ton13}.
The data packets participate in the coding procedure
progressively, and hence the receivers are able to decode packets progressively, leading to a lower decoding delay. However, when the traffic load $\rho$ is high (close to $1$), the average delay increases dramatically on the order of $\Theta\left(\frac{1}{(1-\rho)^2}\right)$ \cite{MWNC, sundararajan2009feedback}.

\emph{Feedback challenge:} To achieve reliable multicast or to improve the delay performance (e.g., \cite{ho2006random,luby2002,shokrollahi2006raptor,MWNC,kumar2008arq,sundararajan2009feedback,parastoo2010optimal,Ton13}), the transmitter needs to collect channel state information or reception status reports (such as ACKs/NAKs) from the receivers through feedback. Conventionally, the feedback overhead increases linearly with the number of receivers, e.g., \cite{ho2006random,luby2002,shokrollahi2006raptor,kumar2008arq,sundararajan2009feedback,parastoo2010optimal,Ton13}, which becomes a system bottleneck when the multicast sessions have a large number of receivers.

In modern wireless networks, multi-channel communications have become commonplace. For instance, 4G/LTE mobile wireless networks are based on OFDM and OFDMA, where a wide spectrum band is divided into many resource blocks, each with 180kHz \cite{holma2009lte}. The 802.11a standard can have $12$ orthogonal channels in the 5GHz band. The availability of multiple channels provides significant flexibility in designing wireless resource allocation algorithms. 


In order to address the above three fundamental challenges, we develop a multi-channel multicast code design in this paper, which can simultaneously achieve high throughput, low delay\footnote{In this paper, we consider the end-to-end delay (including queueing delay and transmission delay), which is measured from the time a packet arrives at the transmitter to the time that the packet is decoded at the receiver. In comparison, the delay metric considered in e.g., \cite{yang2012throughput,MWNC,parastoo2010optimal,Ton13} only accounts for the transmission delay.}, and low feedback overhead. The contributions of this paper are summarized as follows.

\begin{itemize}

\item We propose Multi-channel Moving Window Codes (MC-MWC). The key idea behind MC-MWC is a simple $\mathsf{Merging}$ strategy, through which multiple multicast sessions can be jointly served by the shared multi-channel resources.

\item {High Throughput:} We focus on a many-user many-channel asymptotic regime. First, we derive an algorithm independent lower bound on the number of channels needed for achieving any non-vanishing throughput as the number of receivers increases. Then, we show that MC-MWC achieves the lower bound in an order sense. Hence, MC-MWC achieves order-optimal throughput in the many-user many-channel asymptotic regime. Furthermore, we prove that the number of channels required by a conventional scheme based on the optimal static channel allocation and capacity-achieving codes is doubly-exponentially larger than that required by MC-MWC.

\item {Low Delay:} Using large deviations theory, we show that the delay of MC-MWC decreases linearly as the number of channels grows, while the delay reduction of conventional channel-allocation based schemes (even incorporating with any coding schemes) is no more than a finite constant. To the best of our knowledge, MC-MWC {\it is the first wireless multicast scheme in which the delay decreases linearly as the number of channels grows, with no loss in the per-channel-throughput.}

\item {Low Feedback Overhead:} By combining MC-MWC with the recently proposed anonymous feedback technique \cite{MWNC,Rate_control}, the total feedback overhead of MC-MWC is a constant, independent of not only the number of receivers in each session but also the number of sessions in the network. 

\item Trace-driven and numerical results are provided to validate the analytical results and show that 1) MC-MWC achieves significant throughput and delay improvements in the practical scenarios even when the number of users (channels) is not large, 2) the implementation complexity of MC-MWC is low in practice.

\end{itemize}


 \ifreport
The rest of this paper is organized as follows. In Section \ref{related_work}, we introduce some related works. In Section \ref{sec:system_model}, the system model is described. In Section \ref{sec:code}, we introduce Multi-channel Moving Window Codes (MC-MWC). The throughput and delay performance of MC-MWC are analyzed in Section \ref{sec:throughput} and Section \ref{sec:delay}, respectively. In Section \ref{sec:experiments}, we evaluate the performance of our method through simulations. Finally, we conclude the paper in Section~\ref{sec:conclusion}.
\else
\fi

\section{Related Work}\label{related_work}
In wireless unicast, exploiting the multi-channel resources has been extensively studied. For example, in one line of work, e.g., \cite{Q_ssg,Ouyang:2013}, the multi-channel resources are allocated based on the queue lengths of the users such that throughput optimality is achieved. However, throughput optimality does not necessarily imply good delay performance. Thus, another line of work, e.g., \cite{Bo_optimal,Bo_greedy} proposed delay-based scheduling policies, which achieve optimal throughput and provably good delay performance for multi-channel wireless unicast.

In wireless multicast, multiple receivers may receive the same transmitted packet, and as a result, the work-conservation principle assumed in wireless unicast is violated \cite{Sinha:2016}. In addition, rateless codes or network codes, e.g., \cite{ho2006random,luby2002,shokrollahi2006raptor,MWNC,kumar2008arq,sundararajan2009feedback,parastoo2010optimal,Ton13}. are typically used to achieve good throughput performance. With coding, the receivers need additional time to decode the packets and thus the queueing delay considered in wireless unicast does not fully capture the overall delay. For the above two reasons, the methodologies developed for multi-channel wireless unicast are not applicable to the multicast scenario.

Existing studies about resource allocation in wireless multicast are either channel-statistics based or channel-state based. In the first category, e.g., \cite{Pacifier,lin2013multicast}, it is assumed that the transmitter has access to the channel statistics of the receivers. To realize cooperative multicast, the authors in \cite{Pacifier} proposed a network-coding based multicast scheme, in which the scheduling policy depends on the channel statistics of the receivers. In \cite{lin2013multicast}, to maximize the long-term throughput, the channels are allocated statically according to the channel statistics. The second category, e.g., \cite{Sinha:2016,parastoo2010optimal} assumes that the transmitter has perfect knowledge of the channel state information of the receivers and exploits the opportunistic gain in throughput or delay. A throughput-optimal policy was developed in \cite{Sinha:2016} for wireless multicast with dynamic topology. In \cite{parastoo2010optimal}, the authors proposed an instantly decodable network coding scheme which achieves zero decoding delay at the cost of throughput loss. However, collecting channel state information from all receivers incurs a prohibitive large feedback overhead, rendering them impractical when multicast sessions have a large number of receivers. An open question is {\it how do we obtain a linear delay reduction in the number of channels without loss in the per-channel throughput?}


\section{System Model}\label{sec:system_model}

\begin{figure}[t]
\centering
 \includegraphics[width=2.0in]{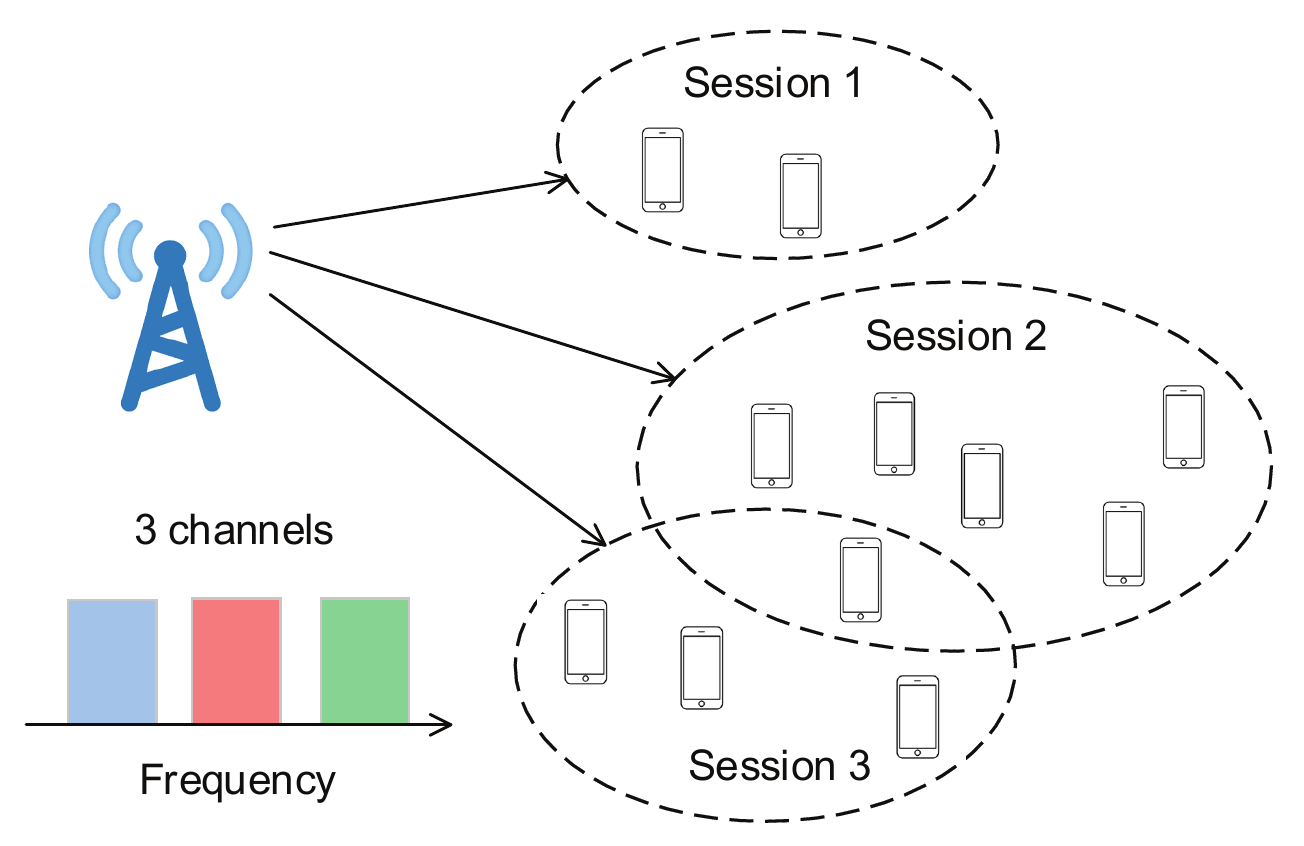}
\vspace{-0.13cm}
\caption{Multi-channel, multi-session wireless multicast.}
\vspace{-0.48cm}
\label{fig:model}
\end{figure}

We consider a multi-channel, multi-session wireless multicast network with $m$ orthogonal channels and $m$ multicast sessions, as shown in Figure~\ref{fig:model}. As in \cite{Q_ssg,Bo_optimal,Bo_greedy}, for ease of presentation, we assume that the number of channels is equal to the number of sessions. Our throughput and delay analysis can be readily generalized to the scenario when the number of channels scales linearly with the number of sessions.

\subsection{Multicast Sessions}
For a multicast session $h\in\{1,\ldots,m\}$, the transmitter needs to send a stream of data packets to a set of receivers, denoted as $\mathcal{S}_h$. For any two different sessions $h$ and $h^\prime$, their sets of receivers $\mathcal{S}_h$ and $\mathcal{S}_{h^\prime}$ are allowed to have arbitrary intersections, i.e., a receiver may be interested in multiple sessions. Let $s_h=|\mathcal{S}_h|\ge 1$ denote the number of receivers in the session $h$. The number of receivers in different sessions may be different and we assume that $s_1,\ldots,s_m$ are {\it i.i.d.} across different sessions\footnote{This assumption can be relaxed such that $\{s_h\}_h$ have heterogeneous distributions. The details are omitted due to space limitations.}. Let $\widehat{s}^{(n)}$ denote a random variable which has the same distribution as $s_h$, where $n$ represents the expected number of receivers in one session, i.e., $n=\mathbb{E}[s_h]$. 

In our model, time is slotted. The packets which need to be served in each session arrive stochastically. Let $a_h[t]\in\mathbb{N}$ denote the number of packet arrivals from session $h$ at the beginning of time-slot $t$. We assume that $\{a_h[t]\}_{t\in\mathbb{N}}$ are stationary and bounded random variables, independent across time-slots, independent of the number of receivers $\{s_h\}_h$. We use $\lambda_h\triangleq \mathbb{E}\left[a_h[t]\right]$ to denote the expected packet arrival rate in session $h$.

\subsection{Channel Model}
The channel between the transmitter and the receivers is assumed to follow the standard broadcast erasure channel model. Let $c_{i,j}[t]$ denote the channel state of receiver $i$ on channel $j$ in time-slot
$t$, given by
\begin{align}
\!\!\!\!c_{i,j}[t]=\left\{
\begin{array}{ll}
1&\text{if the packet sent on channel $j$ can be}\\
&\text{ successfully received by receiver $i$}\\
&\text{in time-slot $t$;}\\
0&\text{otherwise.}
\end{array}\right.\label{cit_def}
\vspace{0.3cm}
\end{align}
As in \cite{MWNC, ho2006random,luby2002,shokrollahi2006raptor,yang2012throughput,kumar2008arq,sundararajan2009feedback,parastoo2010optimal,Ton13}, it is assumed that $c_{i,j}[t]$ are {\it i.i.d.} across different time-slots, and let $\gamma_{i,j}\triangleq\mathbb{P}(C_{i,j}[t]=1)$ denote the probability of successful packet reception at receiver $i$ on channel $j$. We make the following assumption on $\{\gamma_{i,j}\}$.

{\it Assumption (Heterogeneous Channel Conditions):} The channel statistics $\{\gamma_{i,j}\}$ are random variables that have heterogeneous values for different receivers and different channels. Specifically, $\gamma_{i,j}$ are {\it i.i.d.}\footnote{Note that while $\{\gamma_{i,j}\}_{i,j}$ are {\it i.i.d.}, they correspond to the probability of successful packet reception of receiver $i$ on channel $j$. So, these probabilities can be different for different realizations of $\gamma_{i,j}$, hence the channel conditions are heterogeneous.} across receivers and across channels. Let $F_\gamma(\cdot)$ and $\bar{\gamma}\triangleq \mathbb{E}[\gamma_{i,j}]$ denote the the CDF and the expectation of $\gamma_{i,j}$, respectively. Moreover, there exists $\kappa>0$ such that $\lim_{y\to 0^{+}} \frac{F_\gamma(y)}{y}\ge \kappa$.


The above assumptions are more general than the homogeneous network assumptions in \cite{yang2012throughput,kumar2008arq,sundararajan2009feedback,Ton13}, where all $\gamma_{i,j}$ are assumed to be equal and thus fail to reflect the heterogeneous channel conditions in reality. In addition, the existence of $\kappa$ is fairly weak, since it basically says that there is a non-zero probability that $\gamma_{i,j}$ is close to $0$. In Section~\ref{sec:experiments}, we will provide simulation results that are obtained using data traces collected from a software defined radio platform. We will see that the analytical results obtained under the above assumptions are also valid in practice.

\section{Multi-Channel Moving Window Codes}\label{sec:code}
In this section, we propose a Multi-Channel Moving Window Codes (MC-MWC) approach, which is built on the recent work Moving Window Codes (MWC) \cite{MWNC}. Compared with MWC, MC-MWC has two key novelties. First, MWC is designed for the single-channel single-session multicast, and actually we can show that a straightforward generalization of MWC to the multi-channel multi-session multicast setting will lead to poor throughput and poor delay performance. Second, owning to a new $\mathsf{Merging}$ strategy, MC-MWC simultaneously achieves good throughput and delay performance by exploiting the multi-channel resources. In the following, we first introduce a $\mathsf{Merging}$ strategy, and then propose MC-MWC.

\subsection{A Simple Merging Strategy}
The key idea behind MC-MWC is a simple $\mathsf{Merging}$ strategy.

\begin{definition} (Merging)
The $m$ multicast sessions are jointly served by the multi-channel resources as follows: 
\begin{list}
{\bfseries\sffamily Tx's side\,:\hfill}
{
\setlength{\labelwidth}{4em}
\setlength{\labelsep}{0.3em}
\setlength{\leftmargin}{4.3em}
\upshape
}
\item At the transmitter, the streams of packets from all $m$ sessions are merged to form one large multicast session. Then, the transmitter uses the $m$ channels to send the grouped multicast session to all receivers in $\bigcup_{1\le h\le m}\mathcal{S}_h$. 
\end{list}
\begin{list}
{\bfseries\sffamily Rx's side\,:\hfill}
{
\setlength{\labelwidth}{4em}
\setlength{\labelsep}{0.3em}
\setlength{\leftmargin}{4.3em}
\upshape
}
\item In each time-slot, a receiver $i\in\bigcup_{1\le h\le m}\mathcal{S}_h$ listens on all the $m$ channels to receive as many packets as it can. Then, the received packets are used to decode the merged multicast session, including the session(s) receiver $i$ is interested in.
\end{list}
\end{definition} 


Notice that $\mathsf{Merging}$ only describes some high-level multicast transmission rules. The coding, transmission, and feedback procedures of a specific example of $\mathsf{Merging}$ will be explained next in detail. 

In practice, the sessions can be divided into groups and $\mathsf{Merging}$ is applied to the sessions within each group. The choice of group size with $\mathsf{Merging}$ depends on the performance requirements on throughput, delay, and implementation complexity, as will be systematically analyzed and shown in Sections \ref{sec:throughput}, \ref{sec:delay}, and \ref{sec:experiments}.




\subsection{Multi-Channel Moving Window Codes}
\begin{figure}[t]
\centering
\includegraphics[width=3.5in]{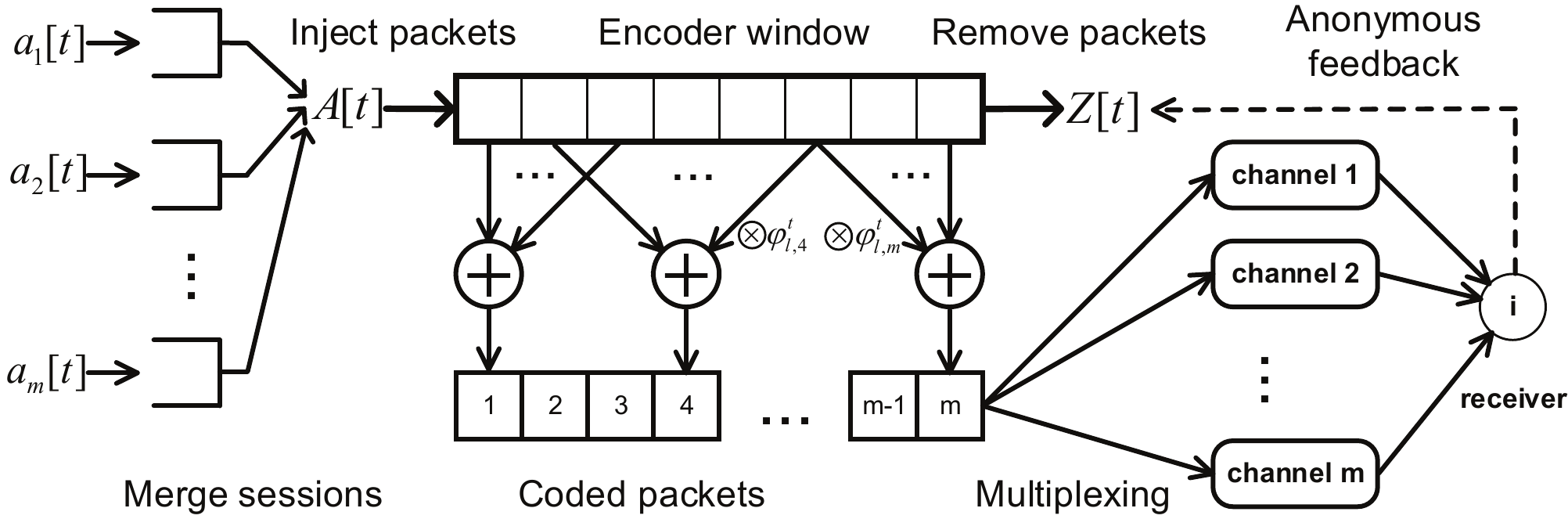}
\vspace{-0.38cm}
\caption{Multi-channel Moving Window Codes.}
\vspace{-0.48cm}
\label{fig:MC-MWC}
\end{figure}

As shown in Figure \ref{fig:MC-MWC}, MC-MWC is comprised of three modules, i.e., the encoder at the transmitter side, the decoder at the receiver side, and the feedback modules.

\subsubsection{Encoder}
At the transmitter, all the packets from all sessions $\{1,\ldots,m\}$ are grouped together to form one large multicast session. The packets in the merged multicast session are indexed as $p_1,p_2,\ldots$, where the packets are indexed according to the order of their arrivals and the tie breaking rule for the packets arriving at the same time-slot is arbitrary. All the newly-arrived packets at the beginning of time-slot $t$ are instantly injected\footnote{The injection rate has to be within the capacity region of MC-MWC, which will be discussed and analyzed in Section~\ref{sec:throughput}.} to an encoder window. At time-slot $t$, the total number of packet arrivals from all the $m$ sessions is denoted as $a[t]\triangleq\sum_{h=1}^m a_h[t]$. Then, the number of packets that the encoder has received up to the beginning of time-slot $t$
is $A[t]$, i.e.,
\begin{align}
A[t]=\sum_{\tau=1}^t a[\tau]=\sum_{\tau=1}^t \sum_{h=1}^m a_h[t].\label{At_def}
\end{align}

To prevent the encoder window from growing indefinitely over time, at the beginning of time-slot $t$, $Z[t]$ packets which are ``determinable''\footnote{``Determinable'' will be formally defined later in the decoder module.} at all receivers $\bigcup_{1\le h\le m}\mathcal{S}_h$ ($Z[t]$ can be determined via a low overhead feedback mechanism described later) are removed from the encoder window. 

At time-slot $t$, the encoder generates $m$ coded packets through linear combinations of the data packets in the encoder window, which are transmitted over the $m$ channels to all receivers $\bigcup_{1\le h\le m}\mathcal{S}_h$. Specifically, the coded packet $x_j[t]$ which is transmitted on the channel $j\in\{1,\ldots,m\}$ in time-slot $t$ is generated by
\begin{align}
x_j[t]=\sum_{l=Z[t]+1}^{
A[t]}\varphi_{l,j}^t\times p_{l},\label{rand_linear_comb}
\end{align}
where $p_l$ denotes the $l^{\text{th}}$ packet of the merged session, ``$\times$'' is the product operator on a Galois field
$GF(2^q)$, and $\left\{\varphi_{l,j}^t\right\}_{\forall l, j, t}$ are independently drawn according to a
uniform distribution on $\{GF(2^q)\}$. The values of
$ A[t]$ and $Z[t]$ are embedded in the packet header of $x_j[t]$. Moreover, $\left\{\varphi_{l,j}^t\right\}_{\forall l, j, t}$ can be known
at each receiver by feeding the same seed to the random number
generators of the transmitter and all the receivers.

\subsubsection{Decoder}

With $\mathsf{Merging}$, a receiver $i\in\bigcup_{1\le h\le m}\mathcal{S}_h$ receives all information sent on the $m$ channels and decode all the $m$ sessions including the session(s) it is interested in, i.e., $\left\{h:i\in\mathcal{S}_h\right\}$.

To facilitate a clear understanding of the decoding procedure, we
restate the definition of {\it ``determinable'' packet} that was originally
defined in \cite{kumar2008arq}.

\begin{definition} (``Determinable'' packet)
A packet $p_l$ is said to be ``determinable'' at a receiver if the receiver has enough information to express $p_l$ as a linear combination of some packets $p_{l+1}, p_{l+2}, \ldots$ with greater indices, and ``indeterminable'' otherwise.
\end{definition}

Let $S_i[t]\in\mathbb{N}$ be the number of ``determinable'' packets at receiver $i$ by the end of time-slot $t$. Define a \emph{virtual
decoder queue}
\begin{align}
Q_i[t]=A[t]-S_i[t]\label{decoder_queue}
\end{align}
for each receiver $i\in\bigcup_{1\le h\le m}\mathcal{S}_h$. Then, $Q_i[t]$ is
the number of packets which have arrived from one of the $m$ sessions but are ``indeterminable'' at receiver $i$ at the end of time-slot $t$.

Suppose at the beginning of time-slot $t$, the packets $p_1,\ldots,p_{l_1}$ are ``determinable'' at receiver $i$ and the total number of packets that the encoder received is $A[t]=l_2$. Similar to \cite{kumar2008arq, MWNC}, when the field size $2^q$ is sufficiently large, for every successfully received packet over the $m$ channels at time-slot $t$, with a high probability, the next ``indeterminable'' packet becomes ``determinable'' until all the packets $p_1,\ldots,p_{l_2}$ are ``determinable''. To better see this, consider the following simple example. The receiver $i$ receives 3 coded packets from channels $2,3,5$ in time-slot $1$, i.e., $x_2[1]=p_1+p_2+2p_3$, $x_3[1]=p_1+2p_2+p_3$ and $x_5[1]=2p_1+p_2+p_3$. From the first packet, $p_1=x_2[1]-p_2-2p_3$, thus, by definition, $p_1$ becomes ``determinable''. Similarly, from the second and third packets, we have $p_2=x_3[1]-x_2[1]+p_3$ and $4p_3=3x_2[1]-x_3[1]-x_5[1]$, by which $p_2$ and $p_3$ are ``determinable'' in turn. Therefore, the evolution of $S_i[t]$ is given by

\begin{eqnarray}
S_i[t]=S_i[t-1]+\min\left\{\sum_{j=1}^m c_{i,j}[t], A[t]-S_i[t-1]\right\}.\label{S_def}
\end{eqnarray}

Combining Equations~\eqref{decoder_queue}, \eqref{S_def} and letting $c_i[t]\triangleq\sum_{j=1}^m c_{i,j}[t]$, the evolution of $Q_i[t]$ can be expressed as:
\begin{align}
Q_i[t]&=\left\{Q_i[t-1]+\sum_{h=1}^m a_h[t]-\sum_{j=1}^m c_{i,j}[t]\right\}^{+}\nonumber\\&=\left\{Q_i[t-1]+a[t]-c_i[t]\right\}^{+}.\label{eq:queue_update}
\end{align}

Similar to \cite{kumar2008arq,MWNC}, decoding occurs at receiver $i$ in time-slot $t$ if and only if all the packets, which have arrived from one of the $m$ sessions are ``determinable'' at receiver $i$, i.e.,
\begin{align}\label{eq:decode_cond}
Q_i[t]=0,
\end{align}
when all packets $p_1,\ldots,p_{A[t]}$ could be decoded.

\subsubsection{Feedback}
Recall that to prevent the encoder window from growing indefinitely over time, $Z[t]$ packets are removed from the encoder window at the beginning of time-slot $t$. To achieve reliable multicast, these $Z[t]$ packets must be ``determinable'' at all receivers $\bigcup_{1\le h\le m}\mathcal{S}_h$ at the beginning of time-slot $t$, and hence $Z[t]$ must be controlled based on the status feedback from all the receivers. To avoid the prohibitively large overhead caused by traditional per-receiver feedback mechanisms, we modify the anonymous feedback mechanism proposed in the recent work \cite{MWNC,Rate_control} to ensure reliable multicast reception at all receivers $i\in\bigcup_{1\le h\le m}\mathcal{S}_h$ with a negligible feedback overhead.

Anonymous feedback was devised for a single multicast session \cite{MWNC,Rate_control}, which guarantees reliable multicast with constant feedback overhead, regardless of the number of receivers. However, without $\mathsf{Merging}$, applying anonymous feedback to the multi-session multicast would incur an overhead which increases with the number of sessions. 

In MC-MWC, anonymous feedback can be applied to the merged session. The key idea is to let the receiver(s), for which the oldest $m$ packets in the encoder are not all ``determinable'', send a NAK in a shared feedback channel at the end of each time slot. This feedback is ``anonymous'' in the sense that the transmitter does not differentiate the actual ID(s) of the receiver(s) sending NAK. As long as the transmitter can detect the existence of NAK signal from feedback, it will keep the oldest $m$ packet in the encoder buffer; otherwise, the oldest $m$ packets are removed from the buffer. This mechanism ensures that 
\begin{align}
Z[t]\le \min_{i\in\bigcup_{1\le h\le m}\mathcal{S}_h}S_i[t-1], 
\end{align}
which guarantees reliable multicast. {\it The anonymous feedback in MC-MWC only requires a short sub-slot to detect the existence of NAK in the shared feedback channel. Thus, the total feedback overhead of MC-MWC is {\bf a constant}, not only independent of the number of receivers in each session, but also independent of the number of sessions in the network.} Compared with conventional schemes, e.g., \cite{ho2006random,luby2002,shokrollahi2006raptor,kumar2008arq,sundararajan2009feedback,parastoo2010optimal,Ton13}, the feedback overhead of MC-MWC is $\Theta(mn)$ times smaller. Compared with MWC \cite{MWNC}, the feedback overhead of MC-MWC is $m$ times smaller. Therefore, MC-MWC effectively reduces the feedback overhead when there are a large number of sessions/receivers. In \cite{Rate_control}, it is shown that anonymous feedback can be easily implemented and incurs a low overhead in practice.


\section{High Throughput of MC-MWC}\label{sec:throughput}
In this section, we analyze the throughput performance of MC-MWC. Noting that multicast is most beneficial when there are a large number of receivers, we focus on a scenario when both the number of receivers and the number of channels scale to infinity. We refer to this setting as the {\it many-user many-channel asymptotic regime}. We emphasize that although our analysis is in the asymptotic regime, the results provide important insights on the practical scenario when the number of users (channels) is not large (e.g., only 10-20 users per session), as illustrated in Section~\ref{sec:experiments}. To facilitate the analysis in this regime, we make the following assumption on the distribution of the number of receivers, i.e., $\{\widehat{s}^{(n)}\}_{n\in\mathcal{N}}$. There is a function $z_s(\cdot)$ such that for any $\alpha\ge 0$ and for any $n\in\mathbb{N}$,
\begin{align}
\mathbb{P}\left(\frac{\widehat{s}^{(n)}}{\mathbb{E}[\widehat{s}^{(n)}]}\le \alpha\right)\le z_{s}(\alpha), \label{eq:n_assume}
\end{align}
and
\begin{align}
\lim_{\alpha\to 0} z_{s}(\alpha)=0. \label{eq:n_assume2}
\end{align}
Note that this is a mild assumption. For instance, the assumption is satisfied when $\widehat{s}^{(n)}$ follows common discrete distributions such as geometric distribution, discrete uniform distribution, Poisson distribution, degenerate distribution, etc.

\subsection{Algorithm Independent Lower Bound}
In the single-channel single-session multicast case, the achievable throughput is bottlenecked by the receiver with the worst channel condition. As a result, with the number of receivers increasing, it is more and more likely that there exists one receiver whose channel condition happens to be poor, leading to a vanishing multicast throughput. We are interested in understanding the following fundamental question {\it in a multi-channel, multi-session multicast: how many multi-channel resources are required to achieve a non-vanishing per-session throughput?} We answer the question by deriving an algorithm independent scaling law on the number of channels required to achieve a non-vanishing throughput as the number of receivers increases.

First, given $m,n$ and the channel statistics $\Gamma=\{\gamma_{i,j}\}_{i,j}$, we define the full capacity region as the set of arrival rate vectors that can be stabilized by some multicast scheme that has the instantaneous channel state information (CSI) of all receivers at the transmitter.
\begin{align}\label{eq:capacity}
\Lambda^{(m,n,\Gamma)}_{\text{Full}}=&\left\{\vec{\lambda}^{(m,n,\Gamma)}=\left(\lambda_1^{(m,n,\Gamma)},\ldots,\lambda_m^{(m,n,\Gamma)}\right)\Big|\right.\nonumber\\&\left.\text{$\vec{\lambda}^{(m,n,\Gamma)}$ can be stabilized by some}\right.\nonumber\\& \text{multicast scheme with CSI at {\sffamily Tx}'s side.}\Big\}.
\end{align}
Note that it is well known that without coding, $\Lambda^{(m,n,\Gamma)}_{\text{Full}}$ can be expressed as the convex hull of all feasible schedules. Nevertheless, with the possibility of inter-session coding, it is generally very difficult to explicitly characterize $\Lambda^{(m,n,\Gamma)}_{\text{Full}}$ even when the channel statistics $\Gamma=\{\gamma_{i,j}\}_{i,j}$ is given. Instead, we derive a fundamental property of $\Lambda^{(m,n,\Gamma)}_{\text{Full}}$ in the many-user many-channel asymptotic regime.

For any session $h$, given the channel statistics $\Gamma=\{\gamma_{i,j}\}_{i,j}$, define the maximum achievable throughput of session $h$ in the full capacity region $\Lambda^{(m,n,\Gamma)}_{\text{Full}}$ as
\begin{align}\label{eq:lambda_up}
\overline{\lambda}^{(m,n,\Gamma)}_h\triangleq \max_{\vec{\lambda}^{(m,n,\Gamma)}\in\Lambda^{(m,n,\Gamma)}_{\text{Full}}} \lambda_h^{(m,n,\Gamma)}
\end{align}
Then we have the following theorem regarding $\overline{\lambda}^{(m,n,\Gamma)}_h$.

\begin{theorem}\label{theorem1}
If the number of channels scales slower than logarithmically\footnote{In this paper, $\log(\cdot)$ denotes the natural logarithm.} with the expected number of receivers, i.e., $m=O\left(\left(\log n\right)^{\delta}\right)$ for some $\delta\in(0,1)$, then the achievable throughput of any multicast scheme vanishes as $n\to\infty$:
\begin{align}
\overline{\lambda}^{(m,n,\Gamma)}_h\,\stackrel{P}{\to}\,0,\;\;\;\forall h\in\{1,\ldots,m\},\label{eq:Theo1}
\end{align}
in which $\stackrel{P}{\to}$ denotes ``convergence in probability''.
\end{theorem}

 \ifreport

\begin{proof}[Proof sketch of Theorem \ref{theorem1}] 
The key idea is to define the set of ``bottleneck'' receivers for a session $h$ as
\begin{align}
\mathcal{B}_\lambda^{(m,n,\Gamma)}=\left\{i\in\mathcal{S}_h\Big|\gamma_{i,j}< \frac{\lambda}{m}, \quad\forall 1\le j\le m\right\}.\nonumber
\end{align}
If $\mathcal{B}_\lambda^{(m,n,\Gamma)}\neq \emptyset$, for any receiver $i\in \mathcal{B}_\lambda^{(m,n,\Gamma)}$, with any possible scheduling and coding scheme, we could upper bound the achievable throughput of receiver $i$ by allocating all $m$ channels to serve receiver $i$ at all time-slots, in which case its throughput is upper bounded by the sum of the capacity of all $m$ channels,
\begin{align} \label{eq:bottle_bd_ske}
\sum_{j=1}^m \gamma_{i,j}< \sum_{j=1}^m \frac{\lambda}{m}=\lambda.
\end{align}
Thus, when $\mathcal{B}_\lambda^{(m,n,\Gamma)}\neq \emptyset$, the achievable throughput of session $h$ is also upper bounded by Equation~\eqref{eq:bottle_bd_ske}.
Then, the focus is to show that for any $\lambda>0$ and any $p<1$, we have
\begin{align} 
\lim_{n\to\infty}\mathbb{P}\left(\mathcal{B}_\lambda^{(m,n,\Gamma)}\neq \emptyset\right)\ge p.\nonumber
\end{align}
\end{proof}
We provide the proof details in Appendix~\ref{sec:pftheorem1}.
\fi
\begin{remark}
 {\normalfont
Theorem~\ref{theorem1} suggests that, to achieve any non-vanishing throughput, the number of channels must scale at least logarithmically with the expected number of receivers, i.e., $m=\Omega\left(\log n\right)$.}
\end{remark}

\subsection{Order Throughput Optimality of MC-MWC}

Now we analyze the achievable throughput of MC-MWC. Given the number of receivers in each session $\{s_h\}_{1\le h\le m}$ as well as the channel statistics $\left\{\gamma_{i,j}\right\}_{i,j}$, the maximum achievable throughput of MC-MWC can be easily derived as follows. For a receiver $i$, it receives information on all the $m$ channels, the sum capacity of which is $\sum_{j=1}^m \gamma_{i,j}$. Recall the capacity of multicast is limited by the worst receiver, the maximum sum throughput of all sessions is then given by
\begin{align}
\sum_{h=1}^m \lambda_h\le \min_{i\in \bigcup_{h=1} ^{m}\mathcal{S}_h} \sum_{j=1}^m \gamma_{i,j}. \label{eq:capacity_merge}
\end{align}
On the other hand, according to Equation~\eqref{eq:queue_update}, $\{Q_i[t]\}_t$ is a random walk on $[0,\infty)$, and has a steady state distribution if Equation~\eqref{eq:capacity_merge} holds. Hence, Equation~\eqref{eq:capacity_merge} is achievable by MC-MWC.

Similar to the definition of $\Lambda^{(m,n,\Gamma)}_{\text{Full}}$, we define the capacity region of MC-MWC. 
\begin{align}
\Lambda^{(m,n,\Gamma)}_{\text{MC-MWC}}=&\left\{\vec{\lambda}^{(m,n,\Gamma)}\Big|\text{$\vec{\lambda}^{(m,n,\Gamma)}$ stabilizable by MC-MWC}\right\}.\nonumber
\end{align}

To characterize $\Lambda^{(m,n,\Gamma)}_{\text{MC-MWC}}$, let us define a convex polytope $\mathcal{P}_{\lambda}^{(m)}\in\mathcal{R}^m$ for a given $m$ and any given $\lambda<\bar{\gamma}$ as
\begin{align}\label{eq:polo}
\mathcal{P}_{\lambda}^{(m)}=&\left\{\vec{\lambda}^{(m)}=\left(\lambda_1^{(m)},\ldots,\lambda_m^{(m)}\right)\Big|\right.\nonumber\\&\frac{1}{m}\sum_{h=1}^m\lambda_h^{(m)}\le \lambda,\lambda_h^{(m)}\ge 0,\forall h\Big\}.
\end{align}
The polytope $\mathcal{P}_{\lambda}^{(m)}\in\mathcal{R}^m$ contains all rate vectors such that the average rate is no greater than $\lambda$.

Then we derive the following theorem regarding $\Lambda^{(m,n,\Gamma)}_{\text{MC-MWC}}$.

\begin{theorem}\label{theorem2}
If the number of channels scale logarithmically with the expected number of receivers, i.e., $m=\left\lceil\frac{1}{(\bar{\gamma}-\lambda)^2}\log n\right\rceil$ for any $\lambda<\bar{\gamma}$, then the achievable per-session throughput of MC-MWC is lower bounded by $\lambda$ in the following sense:
\begin{align}
\lim_{n\to\infty} \mathbb{P}\left(\mathcal{P}_{\lambda}^{(m)}\subseteq \Lambda^{(m,n,\Gamma)}_{\text{MC-MWC}}\right)=1.\label{eq:Theo2}
\end{align}
\end{theorem}

 \ifreport
\begin{proof}[Proof sketch of Theorem \ref{theorem2}] 
	Let us prove that, with MC-MWC, for any $p\in (0,1)$, if $m=\left\lceil\frac{1}{(\bar{\gamma}-\lambda)^2}\log n\right\rceil$. then $\mathcal{P}_{\lambda}^{(m)}\subseteq \Lambda^{(m,n,\Gamma)}_{\text{MC-MWC}}$ with probability no less than $p$, as $n\to\infty$.
	
	First, based on Equations~\eqref{eq:queue_update} and \eqref{eq:capacity_merge}, we show that, 
	\begin{align}
	\mathbb{P}\left(\mathcal{P}_{\lambda}^{(m)}\subseteq \Lambda^{(m,n,\Gamma)}_{\text{MC-MWC}}\right) =\mathbb{P}\left(\frac{1}{m}\min_{i\in \bigcup_{h=1} ^{m}\mathcal{S}_h} \sum_{j=1}^m \gamma_{i,j}\ge \lambda\right). \label{eq:prob_merge_ske}
	\end{align} 
	
	Note that for any $\tau\ge 1$, it can be shown that
	\begin{align}\label{eq:bd_union_ske}
	\mathbb{P}\left(\Big |\bigcup_{h=1}^{m}\mathcal{S}_h\Big|\ge \tau mn\right)&\le \frac{1}{\tau}.
	\end{align}
	
	Then, using Hoeffding's inequality, we derive a lower bound of Equation~\eqref{eq:prob_merge_ske}.
	\begin{align}\label{eq:prob_merge_2_ske}
	&\mathbb{P}\left(\frac{1}{m}\min_{i\in \bigcup_{h=1} ^{m}\mathcal{S}_h} \sum_{j=1}^m \gamma_{i,j}\ge \lambda\right)\nonumber\\\ge&\left(1-\frac{1}{\tau}\right)\left(1-\ceil*{\tau mn}e^{-2(\bar{\gamma}-\lambda)^2m}\right).
	\end{align}
	
	For any $p\in [0,1)$, let $\tau=\frac{2}{1-p}$ and $m=\left\lceil\frac{1}{(\bar{\gamma}-\lambda)^2}\log n\right\rceil$. Combining with Equations~\eqref{eq:prob_merge_ske} and \eqref{eq:prob_merge_2_ske}, we have
	\begin{align}
	&\mathbb{P}\left(\mathcal{P}_{\lambda}^{(m)}\subseteq \Lambda^{(m,n,\Gamma)}_{\text{MC-MWC}}\right) \ge &\frac{p+1}{2}\left(1-o(1)\right),\nonumber
	\end{align}
	where $o(1)$ converges to $0$ as $n\to\infty$.
	
	Noting that $\frac{p+1}{2}\ge p$ and $p$ can be arbitrarily close to $1$, the proof is complete.
	
\end{proof}
	
We provide the proof details in Appendix~\ref{sec:pftheorem2}.

\fi

\begin{remark}
 {\normalfont
Equation~\eqref{eq:Theo2} suggests that if the number of channels scales logarithmically with the number of receivers, with high probability MC-MWC can stabilize all the rate vectors such that the average rate is less than $\bar{\gamma}$.}
\end{remark}
\begin{remark}
 {\normalfont
Comparing Theorem~\ref{theorem2} with Theorem~\ref{theorem1}, we can see that to achieve a non-vanishing throughput, the number of channels required by MC-MWC achieves the algorithm independent lower bound in an order sense. Hence, MC-MWC achieves order-optimal throughput in the many-user many-channel asymptotic regime.}
\end{remark}

\subsection{Throughput Gain over a Conventional Scheme}

Finally, we compare MC-MWC with a conventional scheme, which can be considered as a straightforward extension of \cite{Pacifier,lin2013multicast} to the multi-channel, multi-session multicast setting. First, let us define capacity-achieving code.
\begin{definition} (Capacity-achieving code)
A coding scheme is said capacity-achieving if it can achieve the capacity of the broadcast erasure channel. 
\end{definition}
The conventional scheme is based on the following static channel allocation.

{\it Static Channel Allocation:}
In the static channel allocation, the transmitter allocates the channels according to the channel statistics $\Gamma=\{\gamma_{i,j}\}_{i,j}$. Let $\vec{g}=(g_1,\ldots,g_m)$: $\{1,2,\ldots,m\}\mapsto\{1,2,\ldots,m\}$ represent a one-to-one mapping from the sessions $\{h\}_{1\le h\le m}$ to the channels $\{j\}_{1\le j\le m}$. The static channel allocation is to find a $\vec{g}=(g_1,\ldots,g_m)$ such that the long-term sum throughput of all sessions is maximized. Note that given a static channel allocation $\vec{g}$, the maximum achievable throughput of any session $h$ is bottlenecked by the receiver with the worst channel condition on the channel $g_h$, i.e., $\min_{i\in\mathcal{S}_h}\gamma_{i,g_h}$. Then, the optimal static channel allocation is formulated by
\begin{align}\label{eq:goal_static}
\max_{\vec{g}=(g_1,\ldots,g_m)}\left(\sum_{h=1}^m\min_{i\in\mathcal{S}_h}\gamma_{i,g_h}\right).
\end{align}
Given the optimal channel allocation $\vec{g}^\ast$ to Equation~\eqref{eq:goal_static}, the transmitter encodes the packets from session $h$ using a capacity-achieving coding scheme, such as rateless codes (e.g., \cite{luby2002,shokrollahi2006raptor}) and some network coding schemes (e.g., \cite{ho2006random,kumar2008arq}), and sends the coded packets over the channel $\vec{g}^\ast_h$.

With the conventional scheme, given $m,n$ and the channel statistics $\Gamma=\{\gamma_{i,j}\}_{i,j}$, there is a maximum achievable throughput for each session $h$, denoted as $\overline{\lambda}_h^{(m,n,\Gamma,\text{static})}$. We derive the following theorem regarding $\overline{\lambda}_h^{(m,n,\Gamma,\text{static})}$.
\begin{theorem}\label{theorem3}
If the number of channels scales slower than exponentially with the expected number of receivers, i.e., $m=o\left(\upsilon^{n}\right)$ for all $\upsilon>1$, then the achievable throughput of the optimal static channel allocation with capacity-achieving codes vanishes as $n\to\infty$, that is,
\begin{align}
\overline{\lambda}_h^{(m,n,\Gamma,\text{static})}\,\stackrel{P}{\to}\,0,\;\;\;\forall h\in\{1,\ldots,m\}. \label{eq:Theo3}
\end{align}
\end{theorem}

 \ifreport
\begin{proof}
See Appendix~\ref{sec:pftheorem3}.
\end{proof}
\fi

\begin{remark}
 {\normalfont
Theorem~\ref{theorem3} suggests that, to achieve a non-vanishing throughput, the number of channels must scale at least exponentially with the expected number of receivers, i.e., there exists some $\upsilon>1$ such that $m=\Omega\left(\upsilon^{n}\right)$.}
\end{remark}

\begin{remark}
 {\normalfont
Comparing Theorem~\ref{theorem3} with Theorem~\ref{theorem2}, it can be observed that to achieve a non-vanishing throughput, the multi-channel resource required by the optimal static channel allocation with capacity-achieving codes is doubly-exponentially larger than that required by MC-MWC. }
\end{remark}

\section{Low Delay of MC-MWC}\label{sec:delay}
In this section, we show that MC-MWC achieves significant delay reduction by exploiting the multi-channel resources. We focus on the uniform traffic scenario when the arrival rates for all sessions are equal, i.e., $\lambda_h=\lambda,\forall h$.

Without loss of generality, we focus on analyzing the delay performance of an arbitrary session $h$. Let $p_l^h$ be the $l^{\text{th}}$ packet in session $h$. The delay of the packet $p_l^h$ with respect to a receiver $i\in\mathcal{S}_h$ is defined as the time between the arrival of the packet at the transmitter to the decoding of $p_l^h$ and all the packets in session $h$ with smaller indices, denoted as $D_{i,l}$.

Then, assuming the system is stationary and ergodic, for receiver $i\in\mathcal{S}_h$, the delay
violation probability that the delay of a packet exceeds a
threshold $k$ is given by
\begin{align}\label{exceed_P_def}
\mathbb{P}(D_i>k)=\lim_{L\to\infty}\frac{1}{L}\sum_{l=1}^{L}{1}_{\{D_{i,l}>k\}}.
\end{align}

We first analyze the delay performance of MC-MWC.

Let the time-slots $t_i^d$ $(d=1,2,\ldots)$ be the decoding moments
of receiver $i$ satisfying \eqref{eq:decode_cond}. Suppose that packet $p_l^h$ arrives at
time-slot $t$, which is between two successive decoding moments
$t_i^{d}<t\le t_i^{d+1}$. Then, packet $p_l^h$ and all packets with smaller indices in session $h$ will be decoded in time-slot $t_i^{d+1}$. The delay of packet $p_l^h$ at the
receiver $i$ is
\begin{align}\label{eq:delay}
D_{i,l}=t_i^{d+1}-t.
\end{align}

The following theorem characterizes the delay experienced at a receiver $i$, which is shown to be independent of the channel statistics of the other receivers and $n$.
\begin{theorem}\label{theorem5}
For a receiver $i\in\mathcal{S}_h$ such that $\sum_{j=1}^m \gamma_{i,j}>m\lambda$, the asymptotic decay rate of the delay violation probability of MC-MWC is
\[-\lim_{k\to\infty}\frac{1}{k}\log \mathbb{P}(D_i>k)=\Phi_i^m,\]
where
\begin{align}\label{Ia_def}
\Phi_i^m\!=\!\sup_{\theta\in\mathbb{R}}\left\{-m\log\mathbb{E}\left(e^{-\theta
a_h[t]}\right)\!-\!\sum_{j=1}^m\log\left(\gamma_{i,j} e^\theta\!+\!1\!-\!\gamma_{i,j}\right)\right\}.
\end{align}
In addition, for any $\lambda<\bar{\gamma}$, with probability $1$ we have
\begin{align}\label{Ia_asym}
\lim_{m\to\infty}\!\frac{\Phi_i^m}{m}\!\!=\!\sup_{\theta\in\mathbb{R}}\!\left\{\!-\!\log\mathbb{E}\left(e^{-\theta
a_h[t]}\right)\!-\!\mathbb{E}\left[\log\left(\gamma_{i,j} e^\theta\!+\!1\!-\!\gamma_{i,j} \right)\!\right]\!\right\}.
\end{align}

\end{theorem}

 \ifreport
\begin{proof}[Proof sketch of Theorem \ref{theorem5}] 
	We focus on receiver $1$. Define $I_d\triangleq t_1^{d+1}-t_1^d$ as the interval between the $d^\text{th}$ decoding moment and the $(d+1)^\text{th}$ decoding moment, which can be expressed as
	\begin{align}
	I_d=&\min\left\{t\geq 1: \sum_{\tau=t_1^d+1}^{t_1^d+t}\left(\sum_{h=1}^m a_h[\tau]-\sum_{j=1}^m c_{1,j}[\tau]\right)\le 0\right\},\nonumber
	\end{align}
	indicating that the decoding process is a renewal process. Noting that analyzing $D_1$ directly is very difficult, in the first step, we connect $\mathbb{P}(D_1>k)$ with $\mathbb{P}\left(I_d>b\right)$.
	
	\begin{lemma}\label{lemma_delay_1_ske}
		$\mathbb{P}(D_1>k)$ is upper and lower bounded by
		\begin{align}
		&\frac{\mathbb{P}\left(I_d>k\right)}{m\lambda\mathbb{E}{\big[}I_d{\big]}}\le\mathbb{P}(D_1>k)\le\frac{k\mathbb{P}\left(I_d>k\right)+\sum_{b=k}^\infty
			\mathbb{P}\left(I_d>b\right)}{\lambda\mathbb{E}{\big[}I_d{\big]}}.\nonumber
		\end{align}
	\end{lemma}
	
	Then, using large deviations theory, we derive the decay rate for $I_d$.
	\begin{lemma}\label{lemma_delay_2_ske}
		The decay rate of the decoding interval in the steady state is given
		by
		\begin{align}
		-\lim_{b\to\infty}\frac{1}{b}\log
		\mathbb{P}\left(I_d>b\right)=\Phi_1^m,\nonumber
		\end{align}
		where $\Phi_1^m$ is the rate function defined in
		Equation~(\ref{Ia_def}).
	\end{lemma}
	
	By combining Lemma~\ref{lemma_delay_1_ske} and Lemma~\ref{lemma_delay_2_ske}, the decay rate of $D_1$ is derived.
	
\end{proof}
We provide the proof details in Appendix~\ref{sec:pftheorem5}.

\fi

\begin{remark}
 {\normalfont
Theorem~\ref{theorem5} suggests that with MC-MWC, $\mathbb{P}(D_i>k)\approx e^{-\Phi_i^m k}$. From Equation~\eqref{Ia_asym}, the delay improvement with MC-MWC is asymptotically linear in the number of channels $m$. Hence, MC-MWC could dramatically reduce delay by leveraging the multi-channel resources.}
\end{remark}


Next, we show a lower bound on the delay performance of a general class of conventional schemes defined as follows. 

\begin{definition} (Channel-allocation based schemes)
A multicast scheme is said to be channel-allocation based, if different sessions are allocated to and served by different channels. The channel allocation can be either static over time (as formulated by Equation~\eqref{eq:goal_static}), or dynamically adapted across time-slots based on the instantaneous channel state information of all the receivers and all time-slots. To reduce the technical difficulties, it is assumed that one session can be allocated with at most one channel in each time-slot, which makes sense considering that there are $m$ channels and $m$ sessions with $\lambda_h=\lambda,\forall h$. Given the channel allocation, the transmitter may serve the multicast sessions by employing any possible coding scheme, including but not limited to rateless codes and network codes, e.g., \cite{ho2006random,luby2002,shokrollahi2006raptor,MWNC,kumar2008arq,sundararajan2009feedback,parastoo2010optimal,Ton13}.
\end{definition}

\begin{theorem}\label{theorem6}
When $\mathbb{P}\left(a_h[t]>1\right)>0$, for any receiver $i\in\mathcal{S}_h$, the decay rate of delay in any channel-allocation based scheme is upper bounded by
\begin{align}\label{eq:I_bd}
-\lim_{k\to\infty}\frac{1}{k}\log \mathbb{P}(D_i>k)\le\Phi^{\text{const}},
\end{align}
where 
\begin{align}
\Phi^{\text{const}}=\sup\left\{\theta>0:\log\mathbb{E}\left(e^{\theta
a_h[t]}\right)<\theta\right\}\nonumber
\end{align}
is a constant independent of the number of channels $m$.
\end{theorem}

 \ifreport
\begin{proof}
See Appendix~\ref{sec:pftheorem6}.
\end{proof}
\fi

\begin{remark}
 {\normalfont
Theorem~\ref{theorem6} suggests that the conventional channel-allocation based schemes (even incorporating any coding schemes), can achieve at most a constant delay improvement (not a function of the number of channels) by exploring the multi-channel resources.}
\end{remark}

\begin{remark}
 {\normalfont
The intuition behind Theorem~\ref{theorem6} is as follows. For a queueing system, the delay of packets typically originates from two sources: 1) the stochastic and bursty arrival process; 2) the stochastic service process. Note that with a channel-allocation based scheme, a receiver can receive at most one packet from any given session in one time-slot. Even when the receiver could always receive one packet in each time-slot, there is a lower bound of the delay violation probability which originates from the stochastic and bursty arrival process. The lower bound is independent of the number of channels. However, owing to $\mathsf{Merging}$, MC-MWC is not limited to the assumption that a receiver can receive at most one packet from any given session in one time-slot.}
\end{remark}

\section{Simulations}\label{sec:experiments}
In this section, we provide trace-driven and numerical results to investigate the throughput, delay and implementation complexity performance of MC-MWC.

\subsection{Throughput Performance}
 \ifreport
\begin{figure}[t]
\centering
\includegraphics[width=1.9in]{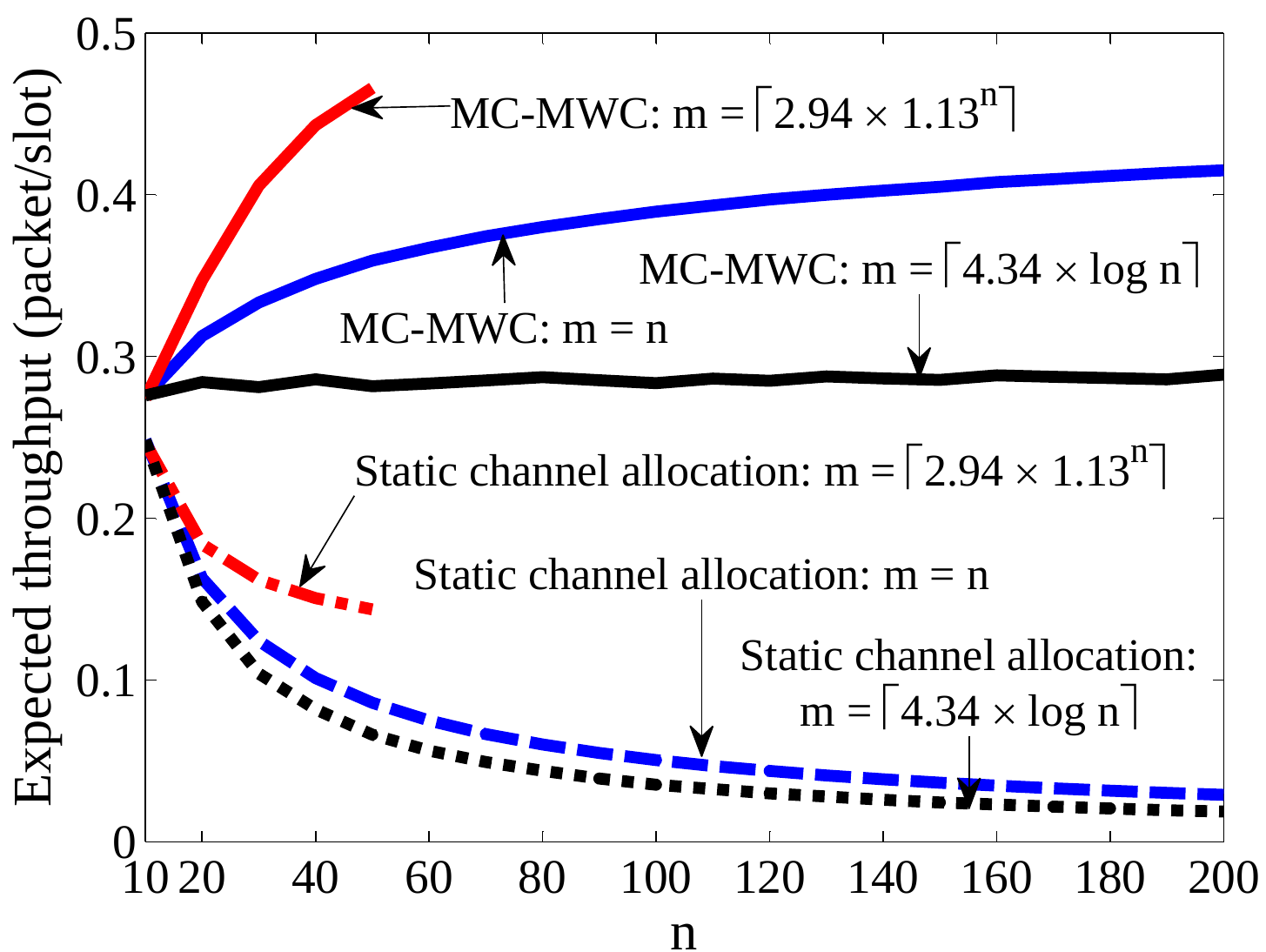}
\vspace{-0.25cm}
\caption{Throughput behavior of MC-MWC vs. static channel allocation with capacity-achieving codes in a heterogeneous network.}
\vspace{-0.25cm}
\label{fig:vs_static}
\end{figure}
To validate our analytical results, we consider a network with heterogeneous channel conditions, where $\gamma_{i,j}$ is uniformly distributed in the interval $[0,1]$. Each of the $m$ multicast sessions has $n$ receivers. We consider the case which is {\it most unfavorable} in terms of throughput for $\mathsf{Merging}$: for any two different sessions $h$ and $h^\prime$, the intersection of the sets of receivers $\mathcal{S}_h$ and $\mathcal{S}_{h^\prime}$ is empty.

We compare the achievable throughput of MC-MWC with that of the optimal static channel allocation with capacity-achieving codes. Figure \ref{fig:vs_static} depicts the average achievable throughput of both schemes in $1000$ randomly generated network scenarios, under different types of scaling of $m$ with respect to $n$. It can be observed that the results match well with Theorem~\ref{theorem1} and Theorem~\ref{theorem3}. When $m=\left\lceil 4.34\times\log n\right\rceil$ and thus increases logarithmically with $n$, MC-MWC achieves a non-diminishing throughput while the throughput of the optimal static channel allocation vanishes as $n$ increases. When $m=\lceil2.94\times 1.13^n\rceil$ and therefore scales exponentially with $n$, the throughput of MC-MWC grows and converges to $\bar{\gamma}=0.5$, and the throughput of the optimal static channel allocation still decreases with $n$ which verifies the prediction in Theorem~\ref{theorem1} that, to achieve a non-diminishing throughput, with the optimal static channel allocation, $m$ has to scale at least exponentially with $n$. When $m=n$ and thus scales linearly with $n$, the throughput of MC-MWC increases while the throughput of the optimal static channel allocation decreases with $n$.
 \else 
 \fi
 
To evaluate the throughput performance of MC-MWC in practice, we collect traces from experiments on a software defined radio platform. We use NI PXIe-1082 platform with NI-5791 RF front end for the experiments. The carrier frequency is set to be 2.5 GHz and the bandwidth used is 40 MHz. Experiments are performed in an indoor lab environment to get the CSI measurements. We note that the effects of multipath is obvious: The SNR difference across this 40 MHz bandwidth can be more than 20 dB. Hence, 
we collect the CSI traces at 100 different client locations and use these traces to emulate multicast receivers. Groups are formed randomly among these receivers.
Packet reception probability is calculated from the subband CSI traces using the method in \cite{zhou2016basic}.

\begin{figure}[t]
\centering
\includegraphics[width=1.9in]{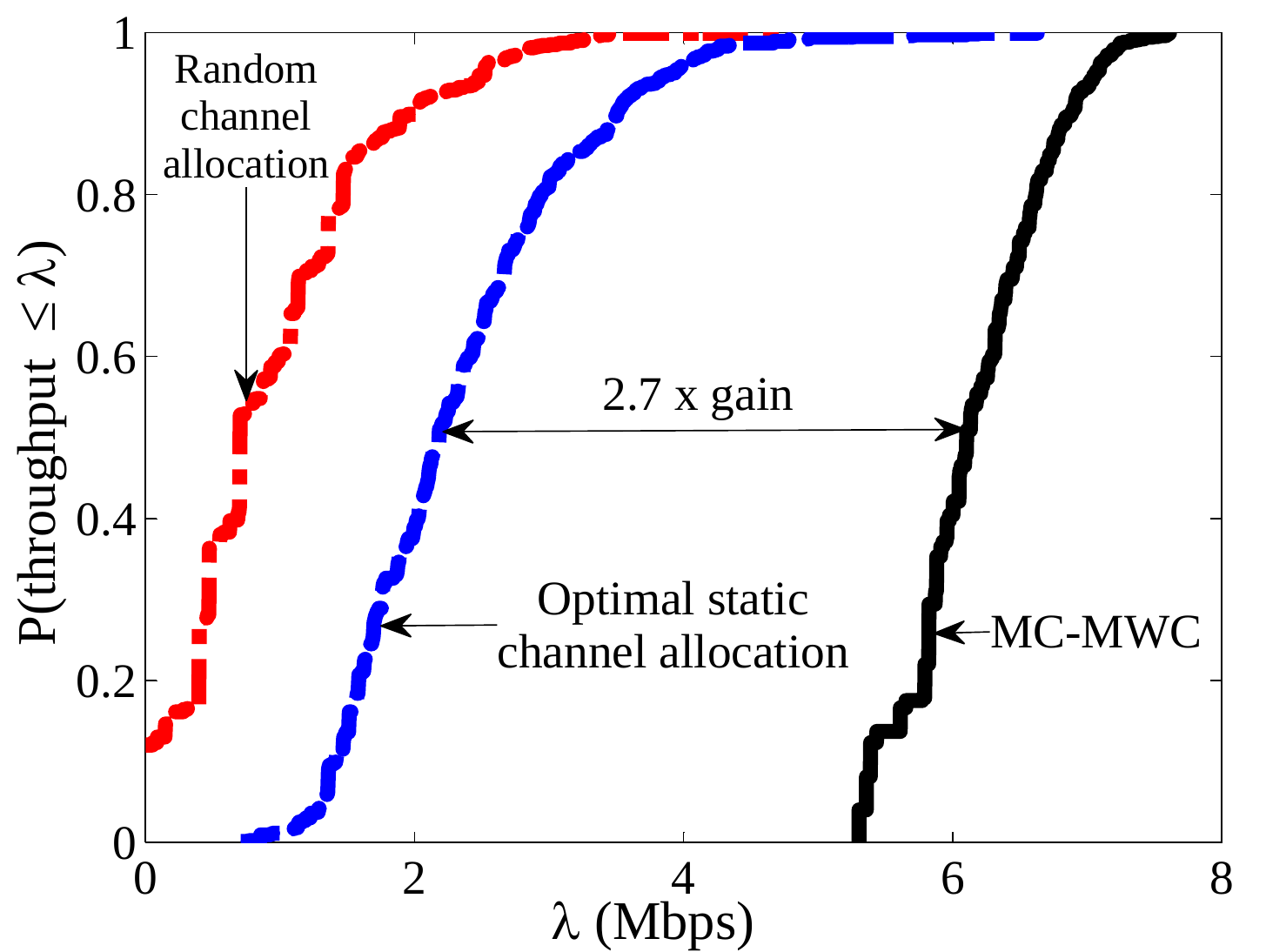}
\vspace{-0.25cm}
\caption{Multicast throughput in trace-driven simulations with $5$ sessions, in which each session has on average $20$ receivers.}
\label{fig:through_cdf}
\end{figure}

Figure \ref{fig:through_cdf} shows the CDF of throughput under different schemes, where $5$ groups are formed with on average $20$ receivers in each group and $1000$ random group formations are performed to get the figure. Here we consider the optimal static channel allocation, as well as a random channel allocation strategy, which assigns channels to sessions randomly. Both the optimal static channel allocation and random channel allocation schemes incorporate random linear network codes (RLNC) \cite{ho2006random}, which is one capacity-achieving code. We can see that MC-MWC shows a 2.7x gain over optimal static channel allocation scheme, which addresses the concern that: {\it in real world, even when the channel conditions are correlated across receivers and channels, and when the number of receivers/channels is not large, the proposed scheme could still have a significant throughput gain.}

\subsection{Delay Performance}

\begin{figure}[t]
\centering
\includegraphics[width=1.9in]{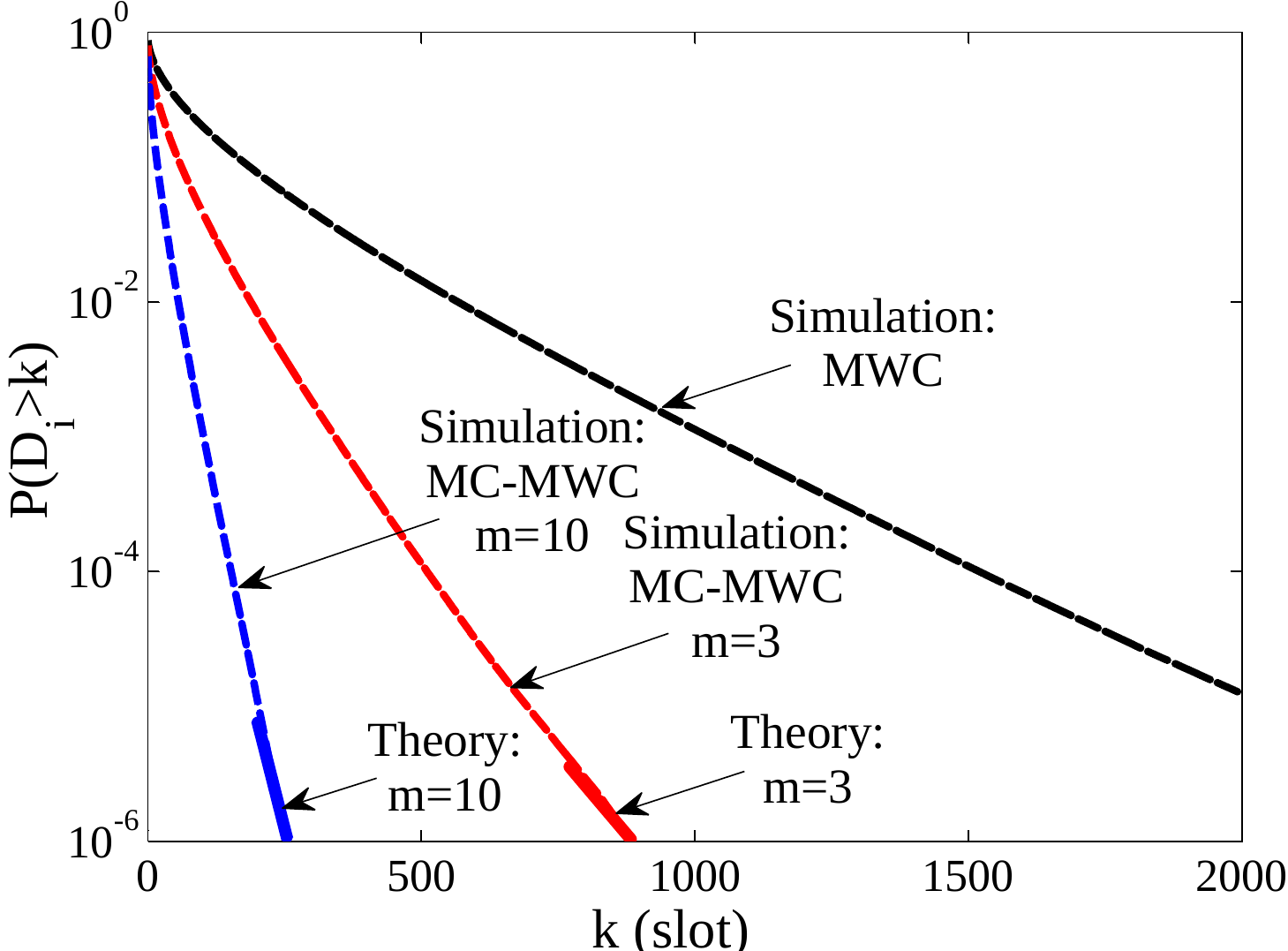}
\vspace{-0.25cm}
\caption{Delay violation probability.}
\label{fig:delay_decay}
\end{figure}

Recall that in the heterogeneous network, MC-MWC achieves a larger throughput than the optimal static channel allocation. To make the delay comparison fair, we consider a homogeneous network scenario, i.e., $\gamma_{i,j}=0.6$ for all receivers and for all channels, in which the maximum achievable throughput of MC-MWC and the optimal static channel allocation are the same. There are $100$ receivers in each session. In each session $h\in\{1,\ldots,m\}$, the packets arrive according to a Bernoulli process with rate $\lambda=0.54$. 

Figure~\ref{fig:delay_decay} plots the delay violation probability of one receiver $i\in\mathcal{S}_h$ with different schemes, i.e., MC-MWC and moving window codes (MWC) \cite{MWNC}. It can be observed that $\mathbb{P}(D_i>k)$ of MC-MWC decays exponentially with $k$ and matches the predicted asymptotic decay rate in Equation~\eqref{Ia_def}. Furthermore, the decay rate is linear in the number of channels $m$, which shows that MC-MWC achieves a significant delay gain over MWC \cite{MWNC} by exploring the multi-channel resources.

\begin{figure}[t]
\centering
\includegraphics[width=1.9in]{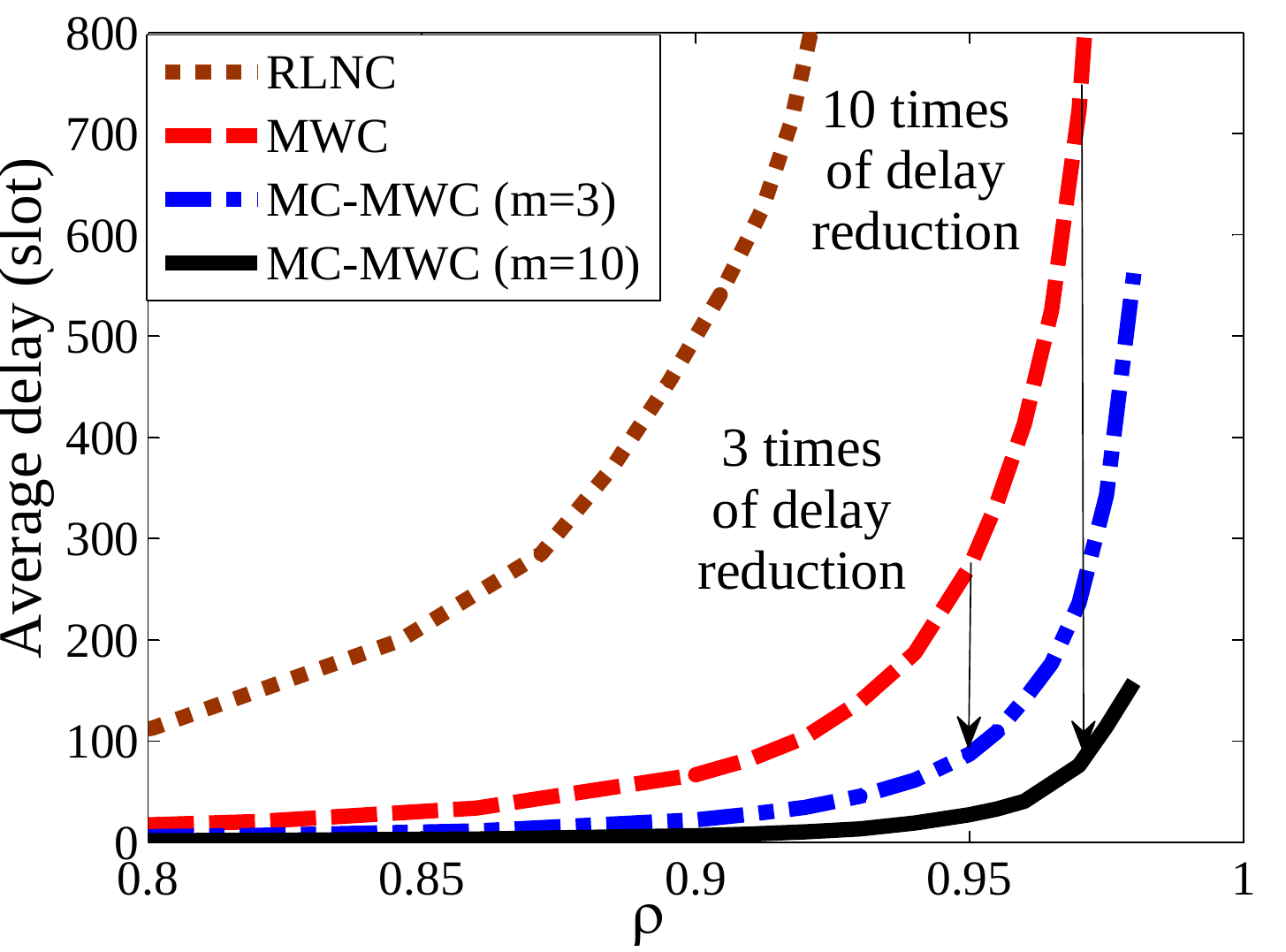}
\vspace{-0.25cm}
\caption{Average delay vs. the traffic load $\rho$.}
\label{fig:av_delay}
\end{figure}

Then, we consider the average delay performance of the same network scenario under different arrival rate $\lambda$. Figure~\ref{fig:av_delay} shows the average delay of different schemes with respect to the traffic load $\rho\triangleq \frac{\lambda}{\gamma_{i,j}}$. From Figure~\ref{fig:av_delay}, we have the following observations. First, the average delay of MC-MWC is much lower than that of RLNC \cite{ho2006random}, which is one of the rateless codes. It is worthy to note that the average delay of other rateless codes, such as LT codes \cite{luby2002} and Raptor codes \cite{shokrollahi2006raptor}, is close to that of RLNC. Second, compared with MWC \cite{MWNC}, MC-MWC achieves a delay reduction which is roughly linear in the number of channels $m$. Third, the delay gain holds for any load $\rho\in(0,1)$.

\subsection{Low Implementation Complexity}
 \ifreport

\begin{figure}[t]
\centering
\includegraphics[width=1.9in]{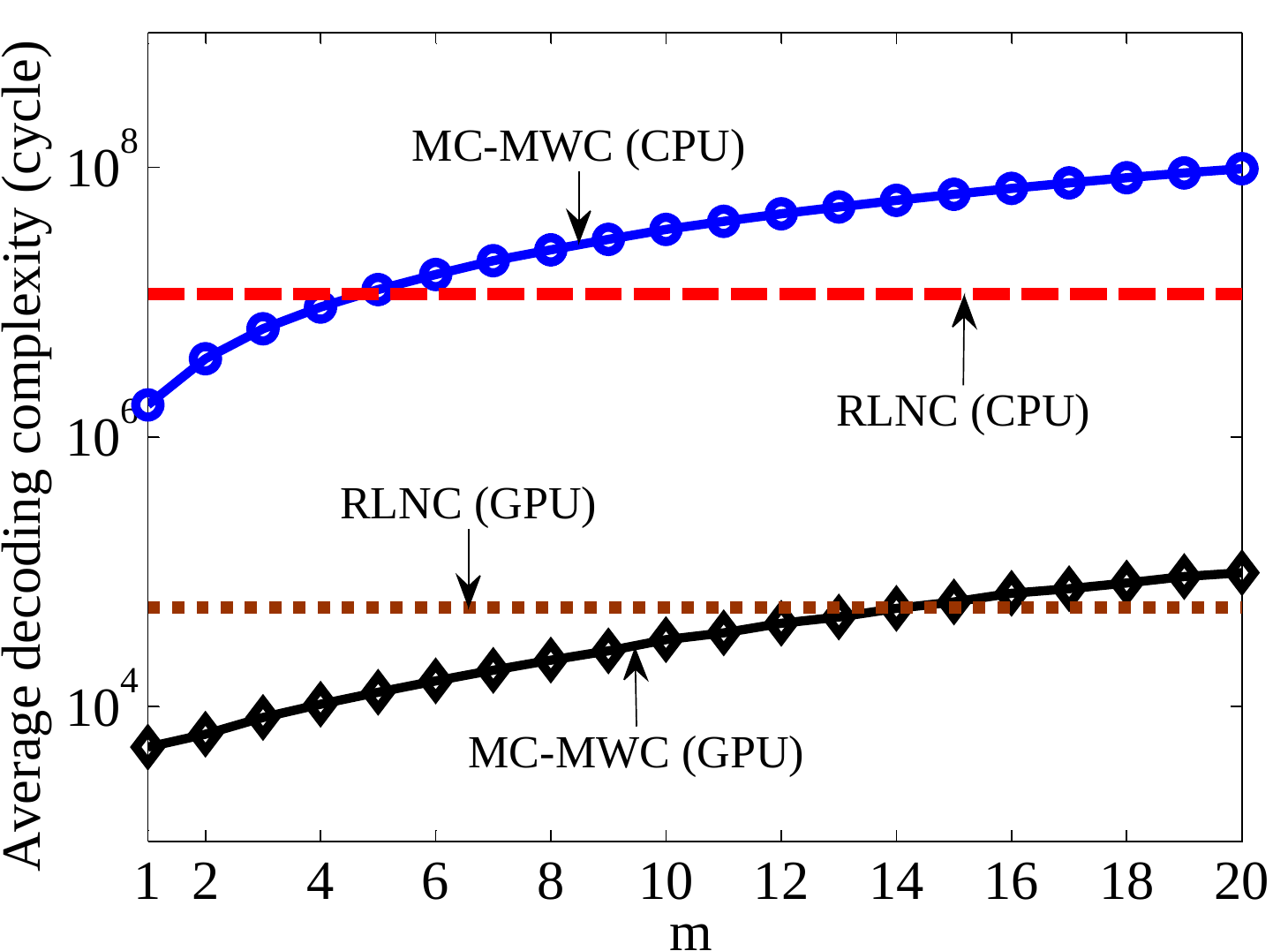}
\vspace{-0.25cm}
\caption{Average decoding complexity on CPU/GPU vs. the number of sessions $m$.}
\vspace{-0.25cm}
\label{fig:av_com}
\end{figure}
With MC-MWC, a receiver needs to decode the sessions it may not be interested in. Hence, it is important to understand the decoding complexity of MC-MWC. To this end, we emulate the decoding procedure of MC-MWC on commercial CPU and GPU, and show that the decoding complexity of MC-MWC is affordable in practice for moderate $m$.

In the emulation, the finite field size in MC-MWC is set to be $256$, and we use a simple table loop-up approach \cite{table} to do the operations on the finite field. Each packet has $1024$ bytes. We emulate the decoding process at a receiver with $\gamma_{i,j}=0.6$ for all channels, where packets arrive according to a Bernoulli process with rate $\lambda=0.48$ such that $\rho=0.8$. We implement two versions of the decoding algorithm, the serial version on a single core CPU (Intel Core i7-2600 CPU clocked at 3.4 GHz), and the parallel version on a GPU (NVIDIA GeForce GTX 860M). Note that the GPU provides 640 shader units clocked at 1029 MHz, and we use 16 blocks and 64 threads per blocks in CUDA. For comparison, we also emulate RLNC \cite{ho2006random} with a block length of $50$ packets, which does not incorporate with $\mathsf{Merging}$.

In Figure \ref{fig:av_com}, we plot the average decoding complexity (in terms of cycles) of different schemes to decode one packet in a session. From Figure \ref{fig:av_com}, we have the following observations. First, as $m$ grows, the decoding complexity of MC-MWC increases on both CPU and GPU, while the decoding complexity of RLNC remains unchanged. This is consistent with our expectation that the decoding complexity with $\mathsf{Merging}$ increases with $m$. Second, for both MC-MWC and RLNC, GPU leverages its ability to decode packets in parallel, and thus dramatically reduce the decoding complexity. Third, despite that MC-MWC requires the receiver to decode $m$ sessions, the decoding complexity of MC-MWC is smaller than that of RLNC when $m$ is small. This is because the moving window coding strategy in general results in a sparser decoding matrix compared with the dense decoding matrix in RLNC, as shown in \cite{MWNC,tassi2016analysis}. Finally, MC-MWC can support $10$ sessions, the throughput of each is more than $100Mbps$ on our GPU. Therefore, the decoding complexity of MC-MWC is low in practice for moderate $m$, in which case MC-MWC already exhibits significant throughput and delay gains as shown previously. 
\fi

\section{Conclusion}\label{sec:conclusion}
In this paper, we develop a Multi-Channel Moving Window Codes (MC-MWC) and prove that it achieves high throughput, low delay, and requires very limited feedback. We verify our theoretical results using trace-driven simulations, and show that the complexity of implementing MC-MWC is low in practice. This new approach, which exploits multi-channel capability, moving window coding, and anonymous feedback, has the potential to finally realize in practice the significant promise of wireless multicast.

\ifreport

{\scriptsize
\bibliographystyle{ieeetr}
\bibliography{reference}} 
\appendices
\section{Proof of theorem \ref{theorem1}}\label{sec:pftheorem1}
Without loss of generality, let us focus on analyzing the achievable throughput of session $1$. 

We want to show that if the number of channels scales a little bit slower than logarithmically with the expected number of receivers, i.e., $m=O\left(\left(\log n\right)^{\delta}\right)$ for some $\delta\in(0,1)$, the achievable throughput of session $1$ diminishes to $0$ for any possible scheme as $n\to\infty$. It is sufficient for us to show that for any $0<\lambda\le 1$ and any $0<p\le 1$, the probability that session $1$ could achieve throughput $\lambda$ is less than $p$ as $n\to\infty$.

Define the set of ``bottleneck'' receivers for session $1$ as
\begin{align}\label{eq:bottle_set}
\mathcal{B}_\lambda^{(m,n,\Gamma)}=\left\{i\in\mathcal{S}_1\Big|\gamma_{i,j}< \frac{\lambda}{m}, \quad\forall 1\le j\le m\right\}.
\end{align}

Notice that if $\mathcal{B}_\lambda^{(m,n,\Gamma)}\neq \emptyset$, for any receiver $i\in \mathcal{B}_\lambda^{(m,n,\Gamma)}$, with any possible scheme, we could upper bound the achievable throughput of receiver $i$ by allocating all $m$ channels to serve receiver $i$ at all time-slots, in which case its throughput is upper bounded by the sum of the capacity of all $m$ channels,
\begin{align} \label{eq:bottle_bd}
\sum_{j=1}^m \gamma_{i,j}< \sum_{j=1}^m \frac{\lambda}{m}=\lambda.
\end{align}
Recall that the throughput of multicast is bottlenecked by its worst receiver. Thus, when $\mathcal{B}_\lambda^{(m,n,\Gamma)}\neq \emptyset$, the achievable throughput of session $1$ is also upper bounded by Equation~\eqref{eq:bottle_bd}.

By the definition in Equation~\eqref{eq:bottle_set}, the probability that a receiver $i\in\mathcal{S}_1$ is not in the ``bottleneck'' set can be given by
\begin{align} \label{eq:p_not_bottle}
\mathbb{P}\left(i\not\in\mathcal{B}_\lambda^{(m,n,\Gamma)}\right)=1-\left(F_\gamma\left(\frac{\lambda}{m}\right)\right)^m,
\end{align}
where $F_\gamma(\cdot)$ is the CDF of $\gamma_{i,j}$.

Subsequently, based on Equation~\ref{eq:p_not_bottle}, we will show that the probability that $\mathcal{B}_\lambda^{(m,n,\Gamma)}\neq \emptyset$ is high when $m=O\left(\left(\log n\right)^{\delta}\right)$ for any $\delta\in(0,1)$.
\begin{align}\label{eq:B_empty}
&\mathbb{P}\left(\mathcal{B}_\lambda^{(m,n,\Gamma)}\neq \emptyset\right)\nonumber\\=&\mathbb{P}\left(\mathcal{B}_\lambda^{(m,n,\Gamma)}\neq \emptyset\big|s_1\le \alpha n\right)\mathbb{P}\left(s_1\le \alpha n\right)+\nonumber\\&\mathbb{P}\left(\mathcal{B}_\lambda^{(m,n,\Gamma)}\neq \emptyset\big|s_1> \alpha n\right)\mathbb{P}\left(s_1> \alpha n\right)\nonumber\\\stackrel{(a)}{\ge}&\left(1-z_s(\alpha)\right)\left(1-\mathbb{P}\left(\mathcal{B}_\lambda^{(m,n,\Gamma)}=\emptyset\big| s_1>\alpha n\right)\right)\nonumber\\\stackrel{(b)}{=}&\left(1-z_s(\alpha)\right)\left(1-\left(1-\left(F_\gamma\left(\frac{\lambda}{m}\right)\right)^m\right)^{s_1}\right)\nonumber\\\stackrel{(c)}{\ge}&\left(1-z_s(\alpha)\right)\left(1-\left(1-\left(F_\gamma\left(\frac{\lambda}{m}\right)\right)^m\right)^{\floor*{\alpha n}}\right)\nonumber\\\stackrel{(d)}{\ge} &\left(1-z_s(\alpha)\right)\left(1-\left(1-\frac{\floor*{\alpha n}\left(F_\gamma\left(\frac{\lambda}{m}\right)\right)^m}{1+\left(\floor*{\alpha n}-1\right)\left(F_\gamma\left(\frac{\lambda}{m}\right)\right)^m}\right)\right),
\end{align}
where step (a) is based on Equations~\eqref{eq:n_assume} and~\eqref{eq:n_assume2} and $\alpha$ is picked such that $z_s(\alpha)<p$, in step (b) Equation~\eqref{eq:p_not_bottle} is applied, step (c) utilizes the condition that $s_1>\alpha n$, and in step (d), we apply the inequality that $(1+x)^r\le 1+\frac{rx}{1-(r-1)x}$ for $x\in[-1,0]$ and $r>1$.

By the assumption that $\lim_{y\to 0^{+}} \frac{F_\gamma(y)}{y}\ge \kappa$, when $m$ is large enough, we have $\left(F_\gamma\left(\frac{\lambda}{m}\right)\right)^m\ge \left(\frac{\kappa\lambda}{2m}\right)^m$. Then it is easy to verify that with $m=O\left(\left(\log n\right)^{\delta}\right)$ for any given $\delta\in(0,1)$, we have
\begin{align} \label{eq:Theo3_limit}
&\lim_{n\to\infty}\floor*{\alpha n}\left(F_\gamma\left(\frac{\lambda}{m}\right)\right)^m\nonumber\\\ge&\lim_{n\to\infty}\floor*{\alpha n}\left(\frac{\kappa\lambda}{2m}\right)^m=\infty.
\end{align}

Inserting Equation~\eqref{eq:Theo3_limit} into Equation~\eqref{eq:B_empty}, we have
\begin{align} \label{eq:B_bd}
\lim_{n\to\infty}\mathbb{P}\left(\mathcal{B}_\lambda^{(m,n,\Gamma)}\neq \emptyset\right)\ge \left(1-z_s(\alpha)\right)\stackrel{(a)}{\ge}1-p,
\end{align}
where step (a) is because $\alpha$ is picked such that $z_s(\alpha)<p$.

Recall that we have shown that when $\mathcal{B}_\lambda^{(m,n,\Gamma)}\neq \emptyset$, the achievable throughput of session $1$ is less than $\lambda$, according to Equation~\eqref{eq:B_bd}, it is impossible for session $1$ to achieve throughput $\lambda$ with probability $p$. Since $\lambda$ and $p$ could be chosen arbitrarily close to $0$, by the Cauchy criterion (see Theorem 6.3.1 in \cite{resnick2013probability}), Equation~\eqref{eq:Theo1} holds, which completes the proof.

\section{Proof of theorem \ref{theorem2}}\label{sec:pftheorem2}


Let us prove that, with MC-MWC, for any $p\in (0,1)$, to guarantee that $\mathcal{P}_{\lambda}^{(m)}\subseteq \Lambda^{(m,n,\Gamma)}_{\text{MC-MWC}}$ with probability no less than $p$, it is sufficient to let the number of channels scale logarithmically with the expected number of receivers, i.e., $m=\left\lceil\frac{1}{(\bar{\gamma}-\lambda)^2}\log n\right\rceil$, as $n\to\infty$.

Recall that with MC-MWC, the maximum sum throughput of all sessions is given by Equation~\eqref{eq:capacity_merge}. On the other hand, according to Equation~\eqref{eq:queue_update}, $\{Q_i[t]\}_t$ is a random walk on $[0,\infty)$, and has a steady state distribution if Equation~\eqref{eq:capacity_merge} holds, which suggests that Equation~\eqref{eq:capacity_merge} is also achievable by MC-MWC. Hence, we have
\begin{align}
\mathbb{P}\left(\mathcal{P}_{\lambda}^{(m)}\subseteq \Lambda^{(m,n,\Gamma)}_{\text{MC-MWC}}\right) =\mathbb{P}\left(\frac{1}{m}\min_{i\in \bigcup_{h=1} ^{m}\mathcal{S}_h} \sum_{j=1}^m \gamma_{i,j}\ge \lambda\right). \label{eq:prob_merge}
\end{align} 

Notice that for any $\tau\ge 1$, we have
\begin{align}\label{eq:bd_union}
\mathbb{P}\left(\Big |\bigcup_{h=1}^{m}\mathcal{S}_h\Big|\ge \tau mn\right)&\stackrel{(a)}{\le}\mathbb{P}\left(\sum_{h=1}^m s_h\ge \tau mn\right)\nonumber\\&\stackrel{(b)}{\le} \frac{\sum_{h=1}^m \mathbb{E}\left[s_h\right]}{\tau mn}= \frac{1}{\tau},
\end{align}
where step (a) uses the fact that $\big|\bigcup_{h=1}^m\mathcal{S}_h\big|\le \sum_{h=1}^m s_h$ due to possible intersections among the sets $\left\{\mathcal{S}_h\right\}_{1\le h\le m}$, and step (b) applies Markov's inequality to the {\it i.i.d.} random variables $\{s_h\}_{1\le h\le m}$.

With Equation~\eqref{eq:bd_union}, we can derive a lower bound of Equation~\eqref{eq:prob_merge}.
\begin{align}\label{eq:prob_merge_2}
&\mathbb{P}\left(\frac{1}{m}\min_{i\in \bigcup_{h=1} ^{m}\mathcal{S}_h} \sum_{j=1}^m \gamma_{i,j}\ge \lambda\right)\nonumber\\\ge & \mathbb{P}\left(\frac{1}{m}\min_{i\in \bigcup_{h=1} ^{m}\mathcal{S}_h} \sum_{j=1}^m \gamma_{i,j}\ge \lambda\Bigg| \Big |\bigcup_{h=1}^{m}\mathcal{S}_h\Big|< \tau mn\right)\times\nonumber\\&\mathbb{P}\left(\Big |\bigcup_{h=1}^{m}\mathcal{S}_h\Big|< \tau mn\right)\nonumber\\\stackrel{(a)}{\ge} &\left(1-\frac{1}{\tau}\right)\mathbb{P}\left(\min_{i\in \bigcup_{h=1} ^{m}\mathcal{S}_h} \sum_{j=1}^m \gamma_{i,j}\ge m\lambda\Bigg| \Big |\bigcup_{h=1}^{m}\mathcal{S}_h\Big|< \tau mn\right)\nonumber\\\ge&\left(1-\frac{1}{\tau}\right)\mathbb{P}\left[\sum_{j=1}^m \gamma_{i,j}\ge m\lambda\right]^{\ceil*{\tau mn}}\nonumber\\\stackrel{(b)}{\ge}&\left(1-\frac{1}{\tau}\right)\left(1-e^{-2(\bar{\gamma}-\lambda)^2m}\right)^{\ceil*{\tau mn}}\nonumber\\\stackrel{(c)}{\ge}&\left(1-\frac{1}{\tau}\right)\left(1-\ceil*{\tau mn}e^{-2(\bar{\gamma}-\lambda)^2m}\right),
\end{align}
where in step (a), Equation~\eqref{eq:bd_union} is applied, in step (b), Hoeffding's inequality is applied on the {\it i.i.d.} random variables $\{\gamma_{i,j}\}_{j=1,\ldots,m}$ which are bounded by the interval $[0,1]$, and step (c) utilizes the inequality that $(1+x)^r\ge1+xr$ for $x\ge -1$ and $r>1$.

Let $\tau=\frac{2}{1-p}$ and $m=\left\lceil\frac{1}{(\bar{\gamma}-\lambda)^2}\log n\right\rceil$. Combining with Equation~\eqref{eq:prob_merge_2}, we have
\begin{align}\label{eq:prob_merge_3}
&\mathbb{P}\left(\frac{1}{m}\min_{i\in \bigcup_{h=1} ^{m}\mathcal{S}_h} \sum_{j=1}^m \gamma_{i,j}\ge \lambda\right)\ge &\frac{p+1}{2}\left(1-o(1)\right),
\end{align}
where $o(1)$ converges to $0$ as $n\to\infty$.

Noting that $\frac{p+1}{2}\ge p$ and $p$ can be arbitrarily close to $1$, by Equation~\eqref{eq:prob_merge_3} and Equation~\eqref{eq:prob_merge}, Equation~\eqref{eq:Theo2} holds. The proof is complete.

\section{Proof of theorem \ref{theorem3}}\label{sec:pftheorem3}
Without loss of generality, we focus on the analysis of the throughput of session $1$.

Now let us prove that if the number of channels scales slower than exponentially with the expected number of receivers, i.e., $m=o\left(\upsilon^{n}\right)$ for any $\upsilon>1$, then the achievable throughput of session $1$ with the optimal static channel allocation vanishes as $n\to\infty$. Note that it is sufficient for us to prove that with the optimal static channel allocation, to achieve any throughput $\lambda\in(0,\bar{\gamma})$ with any probability $p>0$ as $n\to\infty$, the number of channels must scale at least exponentially with the expected number of receivers, i.e., there exists some $\upsilon>1$ such that $m=\Omega\left(\upsilon^{n}\right)$.

Recall that given $\{\gamma_{i,j}\}_{i\in\mathcal{S}_h,j=1,\ldots,m}$, the optimal static channel allocation maximizes Equation~\eqref{eq:goal_static}. By Equation~\eqref{eq:goal_static}, we can upper bound the throughput of session $1$ by allowing the session to choose any of the $m$ channels, i.e.,
\begin{align}\label{eq:upper_static}
\overline{\lambda}_1^{(m,n,\Gamma,\text{static})}\le \max_{1\le j\le m}\min_{i\in\mathcal{S}_1}\gamma_{i,j}.
\end{align}

Then, the probability that session $1$ could achieve throughput $\lambda$ is upper bounded by
\begin{align}\label{eq:prob_static}
&\mathbb{P}\left(\overline{\lambda}_1^{(m,n,\Gamma,\text{static})}\ge\lambda\right)\nonumber\\\le&\mathbb{P}\left(\max_{1\le j\le m}\min_{i\in\mathcal{S}_1}\gamma_{i,j}\ge\lambda\right)\nonumber\\= &\mathbb{P}\left(\max_{1\le j\le m}\min_{i\in\mathcal{S}_1}\gamma_{i,j}\ge\lambda\Big|s_1\le \alpha n\right)\mathbb{P}\left(s_1\le \alpha n\right)+\nonumber\\&\mathbb{P}\left(\max_{1\le j\le m}\min_{i\in\mathcal{S}_1}\gamma_{i,j}\ge\lambda\Big|s_1> \alpha n\right)\mathbb{P}\left(s_1> \alpha n\right)\nonumber\\\le&\mathbb{P}\left(s_1\le \alpha n\right)+\mathbb{P}\left(\max_{1\le j\le m}\min_{i\in\mathcal{S}_1}\gamma_{i,j}\ge\lambda\Big|s_1> \alpha n\right)\nonumber\\\stackrel{(a)}{\le}&z_s(\alpha)+\mathbb{P}\left(\max_{1\le j\le m}\min_{i\in\mathcal{S}_1}\gamma_{i,j}\ge\lambda\Big|s_1>\alpha n\right),
\end{align}
where step (a) uses Equation~\eqref{eq:n_assume}.

Since by our assumption $\{\gamma_{i,j}\}$ are {\it i.i.d.} across different receivers and different channels, then $\{\min_{i\in\mathcal{S}_1}\gamma_{i,j}\}_{1\le j\le m}$ are also {\it i.i.d} when $s_1$ is given. Thus, we have
\begin{align}\label{eq:prob_static_part}
\mathbb{P}\left(\max_{1\le j\le m}\min_{i\in\mathcal{S}_1}\gamma_{i,j}\ge\lambda\Big|s_1\right)&=1-\mathbb{P}\left(\min_{i\in\mathcal{S}_1}\gamma_{i,j}<\lambda\Big|s_1\right)^m\nonumber\\&=1\!-\!\left(\!1\!-\!\mathbb{P}\!\left(\min_{i\in\mathcal{S}_1}\gamma_{i,j}\ge\lambda\Big|s_1\right)\!\!\right)^m\nonumber\\&\stackrel{(a)}{\le} m\mathbb{P}(\gamma_{i,j}\ge \lambda)^{s_1},
\end{align}
where step (a) utilizes the inequality that $(1+x)^r\ge1+xr$ for $x\ge -1$ and $r>1$.

Noticing that the right hand side of Equation~\eqref{eq:prob_static_part} decreases as $s_1$ increases, we could further have
\begin{align}\label{eq:prob_static_part2}
&\mathbb{P}\left(\max_{1\le j\le m}\min_{i\in\mathcal{S}_1}\gamma_{i,j}\ge\lambda\Big|s_1>\alpha n\right)\nonumber\\=&\frac{\mathbb{P}\left(\max_{1\le j\le m}\min_{i\in\mathcal{S}_1}\gamma_{i,j}\ge\lambda, s_1>\alpha n\right)}{\mathbb{P}\left(s_1>\alpha n\right)}\nonumber\\=&\frac{\sum_{s_1=\lceil \alpha n\rceil}^\infty\left\{\mathbb{P}\left(\max_{1\le j\le m}\min_{i\in\mathcal{S}_1}\gamma_{i,j}\ge\lambda\Big| s_1\right)\mathbb{P}\left(s_1\right)\right\}}{\sum_{s_1=\lceil \alpha n\rceil}^\infty\mathbb{P}\left(s_1\right)}\nonumber\\\le &\max_{s_1\ge\lceil \alpha n\rceil} \mathbb{P}\left(\max_{1\le j\le m}\min_{i\in\mathcal{S}_1}\gamma_{i,j}\ge\lambda\Big| s_1\right)\nonumber\\\stackrel{(a)}{\le}&m\mathbb{P}(\gamma_{i,j}\ge \lambda)^{\floor*{ \alpha n}},
\end{align}
where step (a) applies Equation~\eqref{eq:prob_static_part} for $s_1\ge\lfloor\alpha n\rfloor$.

Inserting Equation~\eqref{eq:prob_static_part2} into Equation~\eqref{eq:prob_static}, we have
\begin{align}\label{eq:prob_static2}
&\mathbb{P}\left(\overline{\lambda}_1^{(m,n,\Gamma,\text{static})}\ge\lambda\right)\le z_s(\alpha)+m\mathbb{P}(\gamma_{i,j}\ge \lambda)^{\floor*{ \alpha n}}.
\end{align}

By assumption in Equation~\eqref{eq:n_assume2}, we can pick $\alpha$ such that $z_s(\alpha)< p$. Then, to achieve throughput with probability $p$, we must have
\begin{align}\label{eq:prob_static_target}
\mathbb{P}\left(\max_{1\le j\le m}\min_{i\in\mathcal{S}_1}\gamma_{i,j}\ge\lambda\right)\ge p.
\end{align}

Combining Equation~\eqref{eq:prob_static_target} with Equation~\eqref{eq:prob_static2}, we have
\begin{align}
m\ge \left(p-z_s(\alpha)\right)\left(\frac{1}{\mathbb{P}(\gamma_{i,j}\ge \lambda)}\right)^{\floor*{ \alpha n}}.
\end{align}

By the assumption of heterogeneous network that $F_\gamma(\lambda)>0$ for any $\lambda>0$, we have $\mathbb{P}(\gamma_{i,j}\ge \lambda)<1$ for any $\lambda>0$. Therefore, with static channel allocation, $m$ has to scale at least exponentially with $n$. 

Hence, if $m$ scales slower than exponentially with $n$, Equation~\eqref{eq:prob_static_target} does not hold as $n\to\infty$. Since $\lambda$ and $p$ can be arbitrarily close to $0$, by the Cauchy criterion (see Theorem 6.3.1 in \cite{resnick2013probability}), Equation~\eqref{eq:Theo3} holds. Thus, the achievable throughput of session $1$ vanishes. The proof is complete.

\section{Proof of theorem \ref{theorem5}}\label{sec:pftheorem5}
In this subsection, we analyze the delay violation probability that the
delay experienced by a receiver exceeds a given threshold for
MC-MWC with general \emph{i.i.d.} packet arrivals.
Without loss of generality, we focus on the analysis of the decoding
delay of receiver $1$.

Notice that the arrival processes of different sessions are symmetric, i.e., $a_h[t]$ is {\it i.i.d.} across time-slots and sessions, and $\mathsf{Merging}$ combines all sessions into one large session and does not differentiate among sessions. The delay violation probabilities of all sessions are the same. Hence, it is sufficient for us to analyze the delay violation probability of the combined session after $\mathsf{Merging}$.

Let us define $I_d\triangleq t_1^{d+1}-t_1^d$. Since $\{t_1^d\}_d$ is set of the decoding moments of receiver $1$ that satisfies Equation~\eqref{eq:decode_cond}, we know that $I_d$ represents the interval between the $d^\text{th}$ decoding moment and the $(d+1)^\text{th}$ decoding moment. This fact, combined with the evolution of $Q_i[t]$ given by Equation~\eqref{eq:queue_update}, we can expressed $I_d$ as
\begin{align}
I_d=&\min\left\{t\geq 1: \sum_{\tau=t_1^d+1}^{t_1^d+t}\left(a[\tau]-c_1[\tau]\right)\le 0\right\}\nonumber\\=&\min\left\{t\geq 1: \sum_{\tau=t_1^d+1}^{t_1^d+t}\left(\sum_{h=1}^m a_h[\tau]-\sum_{j=1}^m c_{1,j}[\tau]\right)\le 0\right\}.\label{T_power_def}
\end{align}

The above equation indicates that the decoding process is a renewal process. Let $K_d$ denote the number of packets from all $m$ sessions that arrive at the encoder queue between time-slot $t_1^d$ and time-slot $t_1^{d+1}$, then it can be expressed as
\begin{align}\label{K_j_def}
K_d=\sum_{t=t_1^{d}+1}^{t_1^{d}+I_d}a[t]=\sum_{t=t_1^{d}+1}^{t_1^{d}+I_d}\sum_{h=1}^m a_h[t].
\end{align}

Notice from Equations~\eqref{T_power_def} and \eqref {K_j_def} that $\{I_d\}_d$ are {\it i.i.d.} and $\{K_d\}_d$ are {\it i.i.d.}, we use $\widehat{I}$ and $\widehat{K}$ to denote random variables which have the same distribution as $I_1$ and $K_1$ respectively.

\begin{lemma}\label{lemma_delay_1}
$\mathbb{P}(D_1>k)$ is upper and lower bounded by
\begin{align}\label{exceed_P_bounds}
&\frac{\mathbb{P}\left(\widehat{I}>k\right)}{m\lambda\mathbb{E}{\big[}\widehat{I}{\big]}}\le\mathbb{P}(D_1>k)\le\frac{k\mathbb{P}\left(\widehat{I}>k\right)+\sum_{b=k}^\infty
\mathbb{P}\left(\widehat{I}>b\right)}{\lambda\mathbb{E}{\big[}\widehat{I}{\big]}}.
\end{align}
\end{lemma}
\begin{remark} The proof of Lemma \ref{lemma_delay_1} is
based on a simple observation. For a given delay threshold $k>0$,
the number of packets decoded after an interval $I_d$ must satisfy
the following conditions. 1) If $ {I}_d\le k$, there is no packets
exceeding the threshold $k$. 2) If $ {I}_d> k$, there are at most $
m{I}_d$ packets which exceed the threshold $k$. 3) If $ {I}_d> k$,
there is at least one packet which exceed the threshold $k$.
\end{remark}
\ifreport
\begin{proof}
See Appendix~\ref{pf_lemma_delay_1}.
\end{proof}
\else The proof of Lemma \ref{lemma_delay_1} is relegated to our full technical report online (please see \cite{tech_report}) due to space limitations. \fi

Lemma \ref{lemma_delay_1} shows the connection between $\mathbb{P}(D_1>k)$
and $\mathbb{P}\left(\widehat{I}>b\right)$. Hence, subsequently we
study the probability that the decoding interval in the steady state
exceeds a certain threshold, i.e.,
$\mathbb{P}\left(\widehat{I}>b\right),b\in\mathbb{N}$.

\begin{lemma}\label{lemma_delay_2}
The decay rate of the decoding interval in the steady state is given
by
\begin{align}\label{interval_rate_final}
-\lim_{b\to\infty}\frac{1}{b}\log
\mathbb{P}\left(\widehat{I}>b\right)=\Phi_1^m,
\end{align}
where $\Phi_1^m$ is the rate function defined in
Equation~(\ref{Ia_def}).
\end{lemma}

\ifreport
\begin{proof}
See Appendix~\ref{pf_lemma_delay_2}.
\end{proof}
\else The proof of Lemma \ref{lemma_delay_2} is relegated to our full technical report online (please see \cite{tech_report}) due to space limitations. \fi

Let us pick $\epsilon\in(0,\Phi_1^m)$. By the definition of decay
rate, we can find $B_{\epsilon}\in\mathbb{N}$, such that $\forall
b\in \mathbb{N}, b\ge B_{\epsilon}$, we have
\begin{align}\label{PTK_bounds}
e^{-b(\Phi_1^m+\epsilon)}<\mathbb{P}\left(\widehat{I}>b\right)<e^{-b(\Phi_1^m-\epsilon)}.
\end{align}

Combining Equations~(\ref{exceed_P_bounds}) and (\ref{PTK_bounds})
yields, for $k$ large enough,
\ifreport
\begin{align}\label{exceed_P_bounds2}
&\mathbb{P}(D_1>k)\le \frac{e^{-k(\Phi_1^m-\epsilon)}}{\lambda
\mathbb{E}{\big[}\widehat{I}{\big]}}\left(k+\frac{1}{1-e^{-(\Phi_1^m-\epsilon)}}\right),\nonumber\\
&\mathbb{P}(D_1>k)\ge \frac{e^{-k(\Phi_1^m+\epsilon)}}{m\lambda
\mathbb{E}{\big[}\widehat{I}{\big]}}.
\end{align}
\else
\begin{align}\label{exceed_P_bounds2}
&\frac{e^{-k(\Phi_1^m+\epsilon)}}{\lambda
\mathbb{E}{\big[}\widehat{I}{\big]}}\le \mathbb{P}(D_1>k)\le
\frac{e^{-k(\Phi_1^m-\epsilon)}}{\lambda
\mathbb{E}{\big[}\widehat{I}{\big]}}\left(k+\frac{1}{1-e^{-(\Phi_1^m-\epsilon)}}\right),\mathbb{P}(D_1>k).
\end{align}
\fi On account of $\lim_{k\to\infty}\frac{\log k}{k}=0$,
Equation~(\ref{exceed_P_bounds2}) leads to
\begin{align}
\Phi_1^m-\epsilon\le-\lim_{k\to\infty}\frac{1}{k}\log
\mathbb{P}(D_1>k)\le \Phi_1^m+\epsilon.\nonumber
\end{align}
Since $\epsilon$ can be arbitrarily close to 0, the decay rate of delay given in Equation~\eqref {Ia_def} is proved.

Lastly, we notice that $\{\gamma_{1,j}\}_{1\le j\le m}$ are {\it i.i.d.} across channels. Hence, by the strong law of large numbers, for any $\lambda<\bar{\gamma}$, with probability $1$ we have $\lim_{m\to\infty}\frac{1}{m}\sum_{j=1}^m \gamma_{i,j}=\bar{\gamma}>\lambda$. Thus, Equation~\eqref{Ia_def} is applicable as we take the limit $m\to\infty$. By applying strong law of large numbers to Equation~\eqref {Ia_def}, we have Equation~\eqref{Ia_asym}, which shows that the decay rate of delay increases linearly with the number of channels. The proof is complete.

\section{Proof of theorem \ref{theorem6}}\label{sec:pftheorem6}
Notice that if a session has only one receiver, then the transmitter can dedicatedly minimize the delay of the single receiver. Thus, to derive the lower bound of the delay performance, we assume that there is only one receiver in session $h$, say $1\in\mathcal{S}_h$. With either the static channel allocation or the genie-aided dynamic channel allocation, the receiver $1\in\mathcal{S}_h$ could successfully receive at most one packet in one time-slot. To lower bound the delay performance, we further assume that the receiver successfully receives one packet in each time-slot.

By the definition of delay, a packet is considered as ``decoded'' only if all packets of smaller indices have been ``decoded''. Thus, packets which arrive earlier must also be served earlier. Since there is only one receiver in the session, coding across packets at the transmitter is not necessary and may only increase the delay of the receiver. This suggests that the delay of the single receiver is minimized when the transmitter sends the packets of session $h$ in a First-in First-out (FIFO) manner. Hence, we derive a lower bound of the delay by considering the corresponding FIFO queue $\widetilde{Q}_1$, in which $a_h[t]$ packets arrive and one packet gets served at each time-slot.

According to Theorem 10.4.1 in \cite{srikant2013communication}, the decay rate of the queue length for $\widetilde{Q}_1$ is given by Equation~\eqref{eq:I_bd}. 
\begin{align}\label{eq:Q_decay}
&-\lim_{k\to\infty}\frac{1}{k}\log \mathbb{P}\left(\widetilde{Q}_1[\infty]>k\right)\nonumber\\=&\sup\left\{\theta>0:\log\mathbb{E}\left(e^{\theta
a_h[t]}\right)<\theta\right\}.
\end{align}
Now we derive a connection between the decay rate of the queue length and the decay rate of the delay for the FIFO queue $\widetilde{Q}_1$. The key step is to construct a mapping: each time-slot $\tau$ at which $\widetilde{Q}[\tau]>k$ is mapped to the $(k+1)^{\text{th}}$ packet in $\widetilde{Q}[\tau]$. Since exactly one packet could be served in $\widetilde{Q}$, different time slots are mapped to different packets. Notice there are $k$ packets ahead of each mapped packet, the delay of each mapped packet is at least $k$. Based on this fact, we have
\begin{align}\label{eq:bridge}
&\mathbb{P}\left(\widetilde{Q}_1[\infty]>k\right)=\lim_{t\to\infty}\frac{1}{t}\sum_{\tau=1}^t{1}_{\left\{\widetilde{Q}_1[\tau]>k\right\}}\nonumber\\\stackrel{(a)}{\le}&\lim_{t\to\infty}\frac{\sum_{1\le l\le \sum_{\tau=1}^t a_h[\tau]}1_{\left\{D_{1,l}\ge k\right\}}}{t}\nonumber\\=&\lim_{t\to\infty}\frac{\sum_{\tau=1}^t a_h[\tau]}{t}\times\frac{\sum_{1\le l\le \sum_{\tau=1}^t a_h[\tau]}1_{\left\{D_{1,l}\ge k\right\}}}{\sum_{\tau=1}^t  a_h[\tau]}\nonumber\\\stackrel{(b)}{=}&\lim_{t\to\infty}\lambda\cdot\frac{\sum_{1\le l\le \sum_{\tau=1}^t a_h[\tau]}1_{\left\{D_{1,l}\ge k\right\}}}{\sum_{\tau=1}^t  a_h[\tau]},
\end{align}
where step (a) is due to the property of the mapping we construct, i.e., each time-slot $\tau$ at which $\widetilde{Q}[\tau]>k$ is mapped to a unique packet the delay of which is at least $k$, and in step (b), strong law of large numbers is applied on the \emph{i.i.d.} random variables $\{a_h[\tau]\}_{\forall \tau}$.

Noticing that $\sum_{1\le l\le \sum_{\tau=1}^t a_h[\tau]}1_{\left\{D_{1,l}\ge k\right\}}$ is the number of packets the delay of which is at least $k$ among the the packets in session $h$ arrived up to time-slot $t$. Thus, by the definition of delay violation probability (given by Equation~(\ref{exceed_P_def})), we have
\begin{align}\label{eq:bridge_2}
\lim_{t\to\infty}\frac{\sum_{1\le l\le \sum_{\tau=1}^t a_h[\tau]}1_{\left\{D_{1,l}\ge k\right\}}}{\sum_{\tau=1}^t a_h[\tau]}=\mathbb{P}\left(D_1\ge k\right),
\end{align}
which, together with Equation~\eqref{eq:bridge}, lead to
\begin{align}\label{eq:bridge_3}
\mathbb{P}\left(\widetilde{Q}_1[\infty]>k\right)\le\lambda\mathbb{P}\left(D_1\ge k\right).
\end{align}

Combining Equations~\eqref{eq:Q_decay} and \eqref{eq:bridge_3}, we can see that for the queue $\widetilde{Q}_1$, the decay rate of the delay is no greater than the decay rate of the queue length. Recall that the delay of $\widetilde{Q}_1$ serves as a lower bound of the delay performance of a receiver with conventional schemes without $\mathsf{Merging}$, Equation~\eqref{eq:I_bd} is proved.

In addition, it is easy to verify that when $\mathbb{P}\left(a_h[t]>1\right)>0$, the upper bounded given in Equation~\eqref{eq:I_bd}, is a constant independent of $m$. The proof is complete.

\ifreport

\section{Proof for Lemma \ref{lemma_delay_1}}\label{pf_lemma_delay_1}

Let ${K}_d^k$ denote the number of packets of the combined session after $\mathsf{Merging}$, with decoding delay
greater than the threshold $k$ for the decoding interval ${I}_d,d\in
\mathbb{N}$. 

By the definition of delay violation probability (given by
Equation~(\ref{exceed_P_def})), the numerator can be expressed as
the sum of the number of packets exceeding the threshold in the
decoding intervals,
\begin{align}\label{exceed_P_def2}
\mathbb{P}(D_1>k)&=\lim_{J\to\infty}\frac{\sum_{d=1}^J
{K}_d^k}{\sum_{d=1}^J {K}_d}\nonumber\\
&=\lim_{J\to\infty}\frac{\sum_{d=1}^{J} {K}_d^k} {\sum_{d=1}^{J}
{I}_d}\times\lim_{J\to\infty}\frac{\sum_{d=1}^{J}
{I}_d}{\sum_{d=1}^{J} {K}_d}.
\end{align}
Subsequently, we show how to derive the properties for the two limit
terms on the right side of Equation~(\ref{exceed_P_def2}).

The second limit term is simple. By
Equation~\eqref{eq:decode_cond}, at a decoding moment $t_1^d$,
all packets up to $ A[t_1^d]$ are decoded by receiver
1. If $t=\sum_{d=1}^J {I}_d$, with Equation~(\ref{At_def}) we have
\begin{align}\label{lim_trivial}
\lim_{J\to\infty}\frac{\sum_{d=1}^{J} {I}_d}{\sum_{d=1}^{J}
{K}_d}=\lim_{t\to\infty}\frac{t}{\sum_{\tau=1}^t
\sum_{h=1}^m a_h[\tau]}\stackrel{(a)}{=}\frac{1}{m\lambda},
\end{align}
where in step (a), strong law of large numbers is applied on
\emph{i.i.d.} random variables $\{a_h[\tau]\}_{\forall \tau, \forall h}$.

To bound the first limit term in Equation~(\ref{exceed_P_def2}), we
observe the following facts for the packets decoded after the
interval $ {I}_d$, which can be seen from
Equations~\eqref{T_power_def} and the definition of $K_d^k$.
\begin{enumerate}
\item If $ {I}_d\le k$, there is no packets exceeding the threshold $k$, i.e., $ {K}_d^k=0$.
\item If $ {I}_d> k$, there are at most $m {I}_d$ packets which exceed the threshold $k$, i.e., $ {K}_d^k\le m {I}_d$.
\item If $ {I}_d> k$, there is at least one packet which exceed the threshold $k$, i.e., $ {K}_d^k\ge 1$.
\end{enumerate}

Thus,
\begin{align}\label{1st_limit_bounds}
\lim_{J\to\infty}\frac{\sum_{d=1}^{J} {K}_d^k} {\sum_{d=1}^{J}
{I}_d}&\le\lim_{J\to\infty}\frac{\sum_{d=1}^{J}1_{\left\{
{I}_d>k\right\}} m{I}_d} {\sum_{j=d}^{J}
{I}_d},\nonumber\\\lim_{J\to\infty}\frac{\sum_{d=1}^{J} {K}_d^k}
{\sum_{d=1}^{J}
{I}_d}&\ge\lim_{J\to\infty}\frac{\sum_{d=1}^{J}1_{\left\{
{I}_d>k\right\}}} {\sum_{d=1}^{J} {I}_d}.
\end{align}

Consider $1_{\left\{ {I}_d>k\right\}} {I}_d$ and $1_{\left\{
{I}_j>k\right\}}$ as the rewards earned in interval $ {I}_d$.
According to the renewal reward theory (see Theorem 11.4
\cite{ccinlar1975exceptional}), we have
\begin{align}\label{exceed_P_renewal}
&\lim_{J\to\infty}\frac{\sum_{d=1}^{J}1_{\left\{{I}_d>k\right\}}
m{I}_d}
{\sum_{d=1}^{J}{I}_d}=\frac{m\mathbb{E}\left[1_{\left\{\widehat{I}>k\right\}}\widehat{I}\right]}{\mathbb{E}\left[\widehat{I}\right]},\nonumber\\
&\lim_{J\to\infty}\frac{\sum_{d=1}^{J}1_{\left\{
{I}_d>k\right\}}}{\sum_{d=1}^{J}
{I}_d}=\frac{\mathbb{E}\left[1_{\left\{\widehat{I}>k\right\}}\right]}{\mathbb{E}\left[\widehat{I}\right]}.
\end{align}
Note that
\begin{align}
&\mathbb{E}\left[1_{\left\{\widehat{I}>k\right\}}\widehat{I}\right]=\sum_{b=k+1}^\infty
b\mathbb{P}\left(\widehat{I}=b\right)\nonumber\\&\;\;\;\;\;\;\;\;\;=k\mathbb{P}\left(\widehat{I}>k\right)+\sum_{b=k}^\infty
\mathbb{P}\left(\widehat{I}>b\right),\nonumber\\
&\mathbb{E}\left[1_{\left\{\widehat{I}>k\right\}}\right]=\mathbb{P}\left(\widehat{I}>k\right),
\end{align}
which, by combining with Equations~(\ref{exceed_P_def2}),
(\ref{lim_trivial}), (\ref{1st_limit_bounds}) and
(\ref{exceed_P_renewal}), completes the proof of
Equation~(\ref{exceed_P_bounds}).

\section{Proof for Lemma \ref{lemma_delay_2}}\label{pf_lemma_delay_2}
Based on Equation~(\ref{T_power_def}),
$\mathbb{P}{\big(}\widehat{I}>b{\big)},b\in\mathbb{N}$ can be expressed as
\begin{align}\label{PTk_def}
&\mathbb{P}\left(\widehat{I}>b\right)=\nonumber\\
&\mathbb{P}
\left(\sum_{\tau=1}^{t}
(c_1[\tau]-{a}[\tau])\le
0,\forall 1\leq t\leq b\right)\nonumber\\&\mathbb{P}
\left(\sum_{\tau=1}^{t}
\left(\sum_{j=1}^m c_{1,j}[\tau]-\sum_{h=1}^m{a}_h[\tau]\right)\le
0,\forall 1\leq t\leq b\right).
\end{align}

Notice that $\{{c}_1[\tau]-{a}[\tau]\}_{
\tau\in\mathbb{N}}$ are \emph{i.i.d.} random variables and
$\mathbb{E}\left[{c}_1[\tau]-{a}[\tau]\right]=\sum_{j=1}^m\gamma_{1,j}-m\lambda>0$.
According to the Cramer's Theorem (see Theorem 2.1.24 in
\cite{dembo2010large}),
\begin{align}\label{cramer}
\mathbb{P}\left(\sum_{\tau=1}^{t}
(\widehat{c}_1[\tau]-\widehat{a}[\tau])\le
0\right)=e^{-t\Phi_1^m+o(t)},
\end{align}
where $\Phi_1^m$ is the rate function defined in
Equation~(\ref{Ia_def}). According to the Ballot's Theorem (see
Theorem 3.3 in \cite{weiss1995large}),
\begin{align}
&\mathbb{P}\left(\sum_{\tau=1}^{t}
\left(\widehat{c}_1[\tau]-\widehat{a}[\tau]\right)\le 0,\forall
1\leq t\leq b\right)\nonumber\\=&e^{-b\Phi_1^m+o(b)},\nonumber
\end{align}
if and only if Equation~(\ref{cramer}) holds. Hence, we get the decay rate regarding the
decoding interval $I$, Equation~(\ref{interval_rate_final}) is
proved.

\else 
\fi

\fi

\end{document}